\documentclass[prx,letterpaper,nobalancelastpage,twocolumn,superscriptaddress,nofootinbib]{revtex4}

\usepackage{graphicx}
\usepackage{amsmath}
\usepackage{amssymb}
\usepackage[english]{babel}
\usepackage{color}
\usepackage[version=4]{mhchem}
\usepackage[hidelinks]{hyperref}
\usepackage{dsfont}

\usepackage{soul}
\newcommand{\abs}[1]{\left| #1 \right|}

\newcommand{\ket}[1]{| #1 \rangle}

\newcommand{\Caltech}{Division of Physics, Mathematics and Astronomy, California Institute of Technology, Pasadena, CA 91125, USA}
\newcommand{\Delaware}{Department of Physics and Astronomy, University of Delaware, Newark, DE 19716, USA}
\newcommand{\JQI}{Joint Quantum Institute, National Institute of Standards and Technology and the University of Maryland,
College Park, Maryland 20742, USA}
\newcommand{\Petersburg}{Petersburg Nuclear Physics Institute of NRC ``Kurchatov Institute'', Gatchina, Leningrad district 188300, Russia}

\renewcommand{\cite}[1]{\mbox{\citep{#1}}}

\begin{document}
\title{Alkaline earth atoms in optical tweezers}
\author{Alexandre Cooper}
\affiliation{\Caltech}
\author{Jacob P. Covey}
\affiliation{\Caltech}
\author{Ivaylo S. Madjarov}
\affiliation{\Caltech}
\author{Sergey G. Porsev}
\affiliation{\Delaware}
\affiliation{\Petersburg}
\author{Marianna S. Safronova}
\affiliation{\Delaware}
\affiliation{\JQI}
\author{Manuel Endres}
\affiliation{\Caltech}

\begin{abstract}
We demonstrate single-shot imaging and narrow-line cooling of individual alkaline earth atoms in optical tweezers; specifically, strontium-88 atoms trapped in $515.2~\text{nm}$ light. We achieve high-fidelity single-atom-resolved imaging by detecting photons from the broad singlet transition while cooling on the narrow intercombination line, and extend this technique to highly uniform two-dimensional arrays of $121$ tweezers. Cooling during imaging is based on a previously unobserved narrow-line Sisyphus mechanism, which we predict to be applicable in a wide variety of experimental situations. Further, we demonstrate optically resolved sideband cooling of a single atom close to the motional ground state of a tweezer. Precise determination of losses during imaging indicate that the branching ratio from $^1$P$_1$ to $^1$D$_2$ is more than a factor of two larger than commonly quoted, a discrepancy also predicted by our \textit{ab initio} calculations. We also measure the differential polarizability of the intercombination line in a $515.2~\text{nm}$ tweezer and achieve a magic-trapping configuration by tuning the tweezer polarization from linear to elliptical. We present calculations, in agreement with our results, which predict a magic crossing for linear polarization at $520(2)~\text{nm}$ and a crossing independent of polarization at $500.65(50)~\text{nm}$.  Our results pave the way for a wide range of novel experimental avenues based on individually controlled alkaline earth atoms in tweezers -- from  fundamental experiments in atomic physics to quantum computing, simulation, and metrology implementations.
\end{abstract}
\maketitle

\begin{figure*}[t!]
	\includegraphics[width=\textwidth]{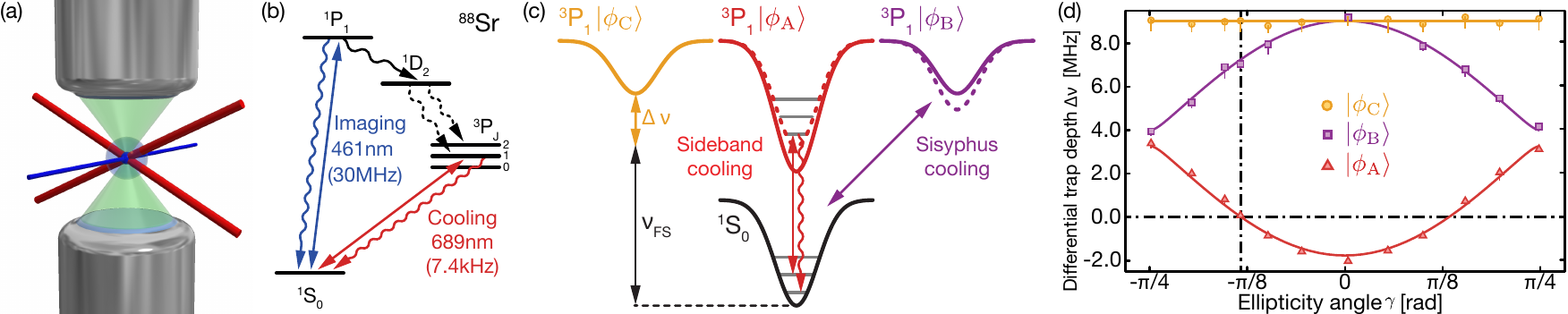}
	\caption{
\textbf{Tweezer trapping of Strontium.}
{(a-b)}~A single strontium atom is trapped in an optical tweezer (propagating upward along the $\hat{z}$ direction) created by focusing a $515.2~\text{nm}$ laser beam through a microscope objective with NA = 0.5 (bottom objective). The atom is imaged by scattering photons on the broad blue transition ($461~\text{nm}$) from a transverse imaging beam, while simultaneously being cooled on the narrow red transition ($689~\text{nm}$) with three red MOT beams (one red beam, overlapped with the imaging beam, is not shown). Fluorescence photons are collected with the bottom objective, while the top objective is mainly used for monitoring the tweezer light.
{(c)}~The applied narrow-line cooling mechanisms, sideband and Sisyphus cooling, depend crucially on the relative trapping potential between ground and three excited sub-levels of $\ce{^{3}P_{1}}$. In a linearly polarized tweezer, these sub-levels can be labeled with angular momentum projection quantum numbers $m_j=-1,0,1$. In elliptical light, rotational symmetry is broken and the sublevels are generally not angular momentum eigenstates anymore. Hence, we label these states with a different notation: $\ket{\phi_\text{C}}$, $\ket{\phi_\text{A}}$, $\ket{\phi_\text{B}}$ (from left to right). Two of the states shift as a function of ellipticity (dashed compared to solid lines). {(d)}~Differential trap depth (proportional to differential polarizability) of the three sublevels of $\ce{^{3}P_{1}}$ as a function of tweezer ellipticity angle $\gamma$, measured with excitation-depletion spectroscopy~(App.~\ref{appendix:tuning} and Sec.~\ref{section:sideband_cooling}). The tweezer polarization is given by $\vec{\epsilon}(\gamma) = \cos{(\gamma)}\hat{x} + i\sin{(\gamma)}\hat{y}$. The solid lines are a fit to the eigenvalues of the AC Stark Hamiltonian (App.~\ref{appendix:tuning}). At the magic ellipticity angle $\tilde{\gamma}=\pm24^\circ$, the differential polarizability between $\ce{^{1}S_{0}}$ and $\ce{^{3}P_{1}}|\phi_\text{A}\rangle$ vanishes. The other two sublevels experience a weaker trapping potential (positive differential trap depths) for all $\gamma$.
}
\label{FigExperimentalSystem}
\end{figure*}

Optical tweezers and related optical micro-potential techniques (OTs) have matured into a powerful tool for quantum science experiments with individually controlled atoms, illustrated by a variety of recent results spanning quantum simulation with Rydberg atoms~\cite{Labuhn2016, Bernien2017, Lienhard2017}, entangling operations~\cite{Browaeys2016, Saffman2016, Levine2018, Kaufman2015}, bottom-up-assembly of Hubbard models~\cite{Kaufman2014, Murmann2015}, and cavity QED implementations~\cite{Thompson2013b, Goban2014, Miller2005}. In these experiments, individual atoms are directly captured from laser-cooled clouds with tweezers or long-wavelength optical lattices~\cite{Frese2000, Schlosser:2001, Nelson:2007}. Some of the more recent technical advances include, e.g., sideband cooling close to the motional ground state in tweezers~\cite{Kaufman2012,Thompson2013}, which has enabled experiments based on coherent collisions~\cite{Kaufman2015} and trapping in the vicinity of nanophotonic structures~\cite{Thompson2013b}. Further, the recently developed atom-by-atom assembly technique~\cite{Barredo2016,Endres2016,Kim2018, Kumar2018, Robens2017} provides means to generate defect-free arrays of currently up to $\sim$60 atoms from initially stochastically loaded OTs~\cite{Schlosser2011, Piotrowicz2013, Nogrette2014, Kaufman2014, Kaufman2015, Nelson:2007}, which has led to the most recent Rydberg quantum simulation applications~\cite{Labuhn2016, Bernien2017, Lienhard2017}. 

In terms of key characteristics, such as effective coherence times, scalability, and controllability, these experiments are now comparable with, and in many ways complementary to, other quantum science platforms with local control, e.g., quantum gas microscopes~\cite{Kuhr2016}, ion traps~\cite{Blatt2012,Monroe2013}, or superconducting qubits~\cite{Devoret2013}. An open question, however, is how distinct properties of non-alkali species can be harnessed for novel and improved implementations in combination with single-atom control via OTs. Of particular interest are alkaline \mbox{earth(-like) atoms (AEAs)}, which offer important features, e.g., narrow and ultra-narrow optical transitions, that have already had a strong impact in various scientific fields, ranging from quantum metrology~\cite{Ludlow2015,Ido2003,Katori2011} and simulation~\cite{Gorshkov2010,Pagano2014,Scazza2014,Cappellini2014,Mancini2015,Kolkowitz2017} to novel approaches for atomic and molecular control~\cite{Stellmer2013, Zelevinsky2006}. 

Here we demonstrate trapping, imaging, and narrow-line cooling of individual AEAs (strontium-88) in an optical tweezer, and extend our imaging technique to highly uniform two-dimensional arrays of 121 tweezers. Our approach builds upon previous experiments for high-resolution imaging of AEAs, including quantum gas microscopes for ytterbium~\cite{Yamamoto2016,Miranda2015} and fluorescence imaging in optical lattice clocks~\cite{Marti2018}.  In addition to resolved sideband cooling, we study a previously unobserved narrow-line Sisyphus cooling mechanism~\cite{Taieb1994, Ivanov2011}, which counteracts fluorescence recoil heating over a wide parameter regime. Interestingly, such single-atom experiments in OTs provide a new tool for determining several important atomic properties of strontium, which we compare to theoretical models. We expect our results to open up an entire spectrum of experiments with individual AEAs controlled with OTs, as described in the outlook. 

\vspace{-3mm}
\section{Tweezer trapping of Strontium}~\label{section:overview}
Tweezer trapping makes use of the AC Stark shift~\cite{Grimm2006}, attracting atoms to the point of maximum intensity in a tightly focused light beam~\cite{Schlosser:2001}. We create a single tweezer, with a beam waist of $w_0\approx500~\text{nm}$, in the center of an ultra-high vacuum cell using a high-resolution objective (Fig.~\ref{FigExperimentalSystem}a, App.~\ref{appendix:experimental_system}). Generating tweezer arrays is discussed in Sec.~\ref{section:tweezer_arrays} and we restrict the discussion to a single tweezer here. To load the tweezer, we overlap it with a laser-cooled cloud of $\ce{^{88}Sr}$ atoms in a narrow-line magneto-optical trap (MOT)~\cite{Katori1999, Loftus2004}. A number of atoms remain in the tweezer after the MOT cloud is dispersed. Subsequently, we induce light-assisted collisions that efficiently removes pairs of atoms~\cite{Schlosser:2001,Jones2006}. As a consequence, the tweezer is filled with at most one atom with an observed occupation probability of $\sim$50~\% (App.~\ref{appendix:parity_projection} and Sec.~\ref{section:fluorescence_imaging}). 

For single-atom detection, we collect blue fluorescence photons while simultaneously applying narrow linewidth cooling to mitigate recoil heating (Fig.~\ref{FigExperimentalSystem}b,c). To this end, we implement a particular type of Sisyphus cooling mechanism~\cite{Taieb1994, Ivanov2011} that relies on the excited state of a narrow optical transition being less trapped than the ground state. In contrast, resolved sideband cooling requires `magic' conditions, i.e., a situation where the ground and excited states experience the same trapping potential~\cite{Ido2003,Ye2008,Leibfried2003}.

In our narrow transition to the $\ce{^{3}P_{1}}$ manifold, we are able to realize both conditions simultaneously for different sublevels, which allows us to study Sisyphus and sideband cooling in a single experimental setting. Specifically, we tune the polarizabilities of the $\ce{^{3}P_{1}}$ sublevels by varying the ellipticity angle $\gamma$ of the tweezer polarization~\cite{Rosenband2018}~(Fig.~\ref{FigExperimentalSystem}d, App.~\ref{appendix:tuning}). For one of these sublevels, we find a `magic-angle' that equalizes ground and excited state polarizability, enabling sideband cooling. The other two sublevels experience significantly weaker trapping for all polarizations, enabling Sisyphus cooling without the need for fine-tuning.

We compare our measurements of differential polarizability at $515.2~\text{nm}$ to theoretical models in App.~\ref{appendix:polarizability}. We find good agreement for the ratio of differential polarizabilities at linear polarization. This quantity provides a new benchmark for theoretical models -- sensitive to even small changes in several matrix elements. Our theoretical models further predict a magic crossing in linearly polarized light at a wavelength of $520(2)~\text{nm}$ and a polarization-insensitive magic crossing at $500.65(50)$ nm. 

\begin{figure}[t!]
	\centering
	\includegraphics[width=\columnwidth]{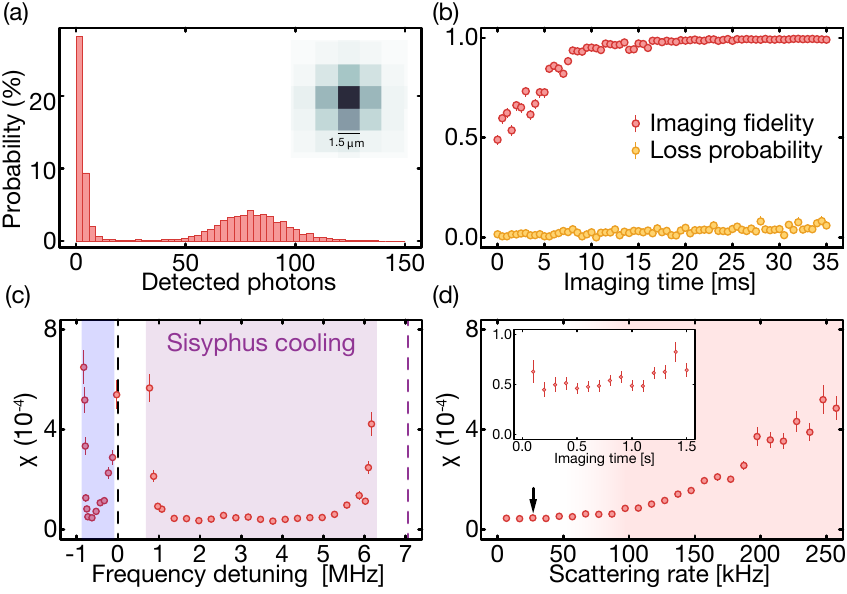}
	\caption{
	\textbf{Imaging in a single tweezer.}
	{(a)}~Histogram of detected photons acquired under typical imaging conditions, showing good discrimination between a zero-atom and single-atom peak. Results are for a single tweezer with magic polarization. Inset: averaged fluorescence image of a single atom (see Sec.~\ref{section:tweezer_arrays} for details).	
	{(b)}~Imaging fidelity and loss probability as a function of imaging time.  Fidelity, defined as the accuracy of image classification, reaches a maximum of $F=99.3(9)~\%$ for sufficiently long imaging times.  However, loss also increases with imaging time. Fidelity is ultimately limited by the estimated number of atoms lost before they can emit enough photons to be detected.  
	{(c)}~The loss coefficient $\chi \equiv -\frac{ln(p_s)}{N}$, where $p_s$ is the survival probability and $N$ is the number of scattered photons, as a function detuning of the cooling light.  A narrow regime of cooling to the red detuned side is interpreted as sideband cooling, while a much broader regime to the blue detuned side is interpreted as Sisyphus cooling. Both regimes achieve the same optimal value of $\chi$.  
	{(d)}~$\chi$ as a function of estimated scattering rate for a fixed imaging time of 200ms. Data shown is for a $1.4~\text{mK}$ trap under Sisyphus cooling. Below $60~\text{kHz}$, $\chi$ approaches a constant minimum value, indicating that losses are dominated by depopulation (white region) and not heating.  As the scattering rate increases beyond $60~\text{kHz}$, cooling can no longer mitigate heating losses (red region).  Inset: $\chi$ versus imaging time, taken at the scattering rate indicated by an arrow ($\sim$27~kHz). $\chi$ stays roughly constant even at very long times.  
	}
	\label{FigFluorescenceImaging}
\end{figure}

\begin{figure*}[t!]
	\centering
	\includegraphics[width=\textwidth]{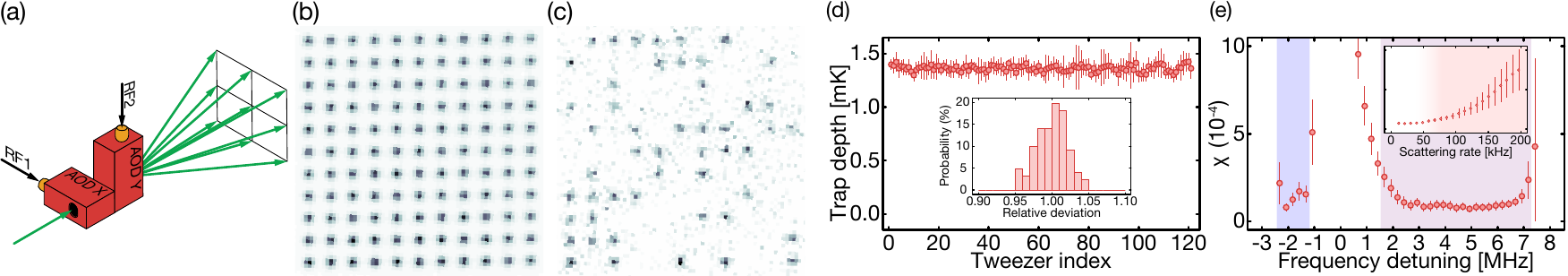}
	\caption{
	\textbf{Tweezer arrays.}
	{(a)}~We create a two-dimensional arrays of tweezers with two perpendicular acousto-optic deflectors (AODs). A 4f-telescope (not shown) maps the light between the two AODs.  Each AOD is driven by a polychromatic RF waveform with tones uniformly spaced in frequency.
	{(b)}~Average fluorescence image (of 6000 experimental runs) of single strontium atoms in a square array of $11\times11$ tweezers. The interatomic distance is $\sim$9$\mu\text{m}$.
	{(c)}~Single-shot image of single strontium atoms in a square array of $11\times11$ tweezers. The filling fraction is close to 0.5.
    {(d)}~Trap depth for all 121 tweezers, as measured by spectroscopy on the $\ce{^{1}S_{0}}\leftrightarrow\ce{^{3}P_{1}}|\phi_\text{C}\rangle$ transition. Inset: a histogram of trap depths across the array. The standard deviation of relative trap depths is $2~\%$, demonstrating homogeneity.
	{(e)}~The loss coefficient $\chi$ as a function of cooling frequency, averaged over an $11\times11$ linearly polarized array. Features are similar to those seen in a single magic tweezer, but pushed further away from the free-space resonance due to larger differential polarizability in linear light. Inset: $\chi$ versus blue scattering rate under Sisyphus cooling, averaged over the array.
	}
	\label{FigTweezerArrays}
\end{figure*}

\vspace{-3mm}
\section{Imaging in a single tweezer}~\label{section:fluorescence_imaging}
Under typical conditions, the observed fluorescence signal on an electron electron multiplying charge-coupled device (EMCCD) camera enables single-shot single-atom resolved detection with high fidelity. Specifically, the histogram of photons detected in a 7$\times$7 box of pixels separates into two resolved peaks of approximately equal area: a zero-atom background peak and a single-atom peak (Fig.~\ref{FigFluorescenceImaging}a). These results are consistent with a single atom occupying the trap in $\sim$50~\% of the repetitions (see also App.~\ref{appendix:parity_projection}). 

We compute a single-shot imaging fidelity $F$ via the \textit{accuracy} of image classification. Images are classified into positives (atom detected) and negatives (no atom detected) by choosing a threshold of detected photons.  The accuracy of classification is defined as the fraction of correctly identified images.  Via an estimate of false positives and false negatives, we estimate this quantity to reach $F=99.3(9)~\%$ in the limit of long imaging times (Fig.~\ref{FigFluorescenceImaging}b, App.~\ref{appendix:fluorescence_imaging}). These values are quoted for a trap depth of 1.4~mK.  We have briefly studied imaging in shallower traps and are able to achieve fidelities higher than $98~\%$ for traps at least as shallow as $300~\mu$K.  

Although we are able to correctly identify the presence or absence of an atom with high fidelity, we find that a small fraction of atoms is lost during the imaging process.  In the histogram, loss manifests itself as a small, roughly flat distribution bridging the single and no-atom peaks. This bridge stems from atoms that are lost before the end of the imaging period and, therefore, result in fewer scattered photons. We emphasize, however, that loss during imaging does not imply that an atom was not detected, as most atoms which are lost still emit enough photons to be above the classification threshold. Nonetheless, the imaging fidelity at long times is ultimately limited by atoms lost before they can emit enough photons to be detected (App.~\ref{appendix:fluorescence_imaging}).  

To quantify loss, we take two consecutive images and define the survival probability $p_s$ of detected atoms as the probability of detecting an atom in the second image conditional on an atom being detected in the first. As loss grows with imaging time, one might want to seek a compromise between fidelity and survival fraction. As typical numbers, we quote $F\sim 99~\%$ at a survival probability of $p_s\sim 97~\%$ for an imaging time of $\sim$20~ms (Fig.~\ref{FigFluorescenceImaging}b).

Under optimized imaging conditions, we find that the experimentally observed survival probability $p_s$ is compatible with an exponential loss in scattered photons, $p_s \approx \exp(- \chi \cdot N)$, where $N$ is an estimator for the number of scattered blue photons (Fig.~\ref{FigFluorescenceImaging}c,d and App.~\ref{appendix:fluorescence_imaging}). For example, we observe that the loss coefficient $\chi$, defined as $\chi \equiv -\frac{ln(p_s)}{N}$, is constant as a function of imaging time during which $N$ grows (inset Fig.~\ref{FigFluorescenceImaging}d). For optimized cooling parameters, we find that $\chi$ is roughly independent of scattering rate for blue scattering rates below $\sim$60~kHz (Fig.~\ref{FigFluorescenceImaging}d). Furthermore, in this limit of low blue scattering rates, we find approximately the same $\chi$ in a wide range of red cooling parameters (Fig.~\ref{FigFluorescenceImaging}c).  

These observations are compatible with a loss mechanism that depopulates the excited state $\ce{^{1}P_{1}}$ via a weak decay channel $\ce{^{1}P_{1}}\rightarrow \ce{^{1}D_{2}}$ (Fig.~\ref{FigExperimentalSystem}b). In our trapping wavelength, $\ce{^{1}D_{2}}$ is strongly anti-confined such that we expect atoms to be ejected faster than they can decay into the triplet manifold. Assuming that all decay into $\ce{^{1}D_{2}}$ results in loss, $\chi^{-1}$ provides a lower bound for the branching ratio between decaying back into $\ce{^{1}S_{0}}$ compared to decaying into $\ce{^{1}D_{2}}$. We find $\chi^{-1}$ to be in the range from $17(3)\times10^3$ to $24(4)\times10^3$ depending on our assumption on the blue emission pattern, which is consistent with an \textit{ab initio} prediction for the branching ratio of $20.5(9)\times10^3$~(App.~\ref{appendix:fluorescence_imaging}). Note in comparison the commonly quoted branching ratio of $50\times10^3$~\cite{Hunter1986}. We discuss strategies for mitigating this depopulation loss in the outlook.

We find the lowest loss coefficients $\chi$ in two distinct red cooling regimes, attributed to sideband and Sisyphus cooling (Fig.~\ref{FigFluorescenceImaging}c). We cool atoms with the $689~\text{nm}$ light simultaneously while driving the blue transition. On the red detuned side of the $689~\text{nm}$ free-space resonance, we observe a narrow cooling feature which we interpret as sideband cooling on the magic-tuned transition to $|\phi_\text{A}\rangle$. On the blue detuned side, where we excite a non-magic transition, there is a much broader feature which we interpret as Sisyphus cooling~(Section~\ref{section:sisyphus_cooling}). The cooling light is provided by three counterpropagating red MOT beams, although we have observed that a single non-counterpropagating beam achieves similar fidelity in the Sisyphus regime, compatible with the interpretation that cooling in this regime is not provided by photon recoil but rather by differential potential energy between ground and excited state. 

\vspace{-3mm}
\section{Tweezer arrays}~\label{section:tweezer_arrays}
We now generalize this imaging strategy to two-dimensional arrays of tweezers. At the same time, this serves as a proof-of-principle for larger-scale two-dimensional tweezer array generation with acousto-optic deflectors (AODs), which have previously been employed for one-dimensional arrays of up to $100$ sites~\cite{Endres2016} and two-dimensional arrays of four~\cite{Lester2015} or 16 sites~\cite{Zimmermann2011}. To this end, we generate a square array of $11\times11=121$ tweezers using two AODs oriented perpendicularly to one another (Fig.~\ref{FigTweezerArrays}a-c), each driven by a polychromatic radiofrequency (RF) signal~(App.~\ref{appendix:experimental_system}). Having shown effective cooling in a magic-tuned tweezer, we choose linear tweezer polarization here instead. This choice aides in maintaining polarization uniformity across the array and lets us explore how cooling features change with modified differential polarizabilities.

We achieve homogeneous trap depths across the array  with a peak to peak variation of $<5~\%$ and a standard deviation of $2~\%$ (Fig.~\ref{FigTweezerArrays}d). To obtain this level of uniformity, we start by coarsely uniformizing the trap depths by imaging the trapping light onto a CMOS camera and feeding back to the amplitudes of the RF tones. Fine uniformization is achieved by spectroscopy on the $\ce{^{1}S_{0}}\leftrightarrow\ce{^{3}P_{1}}|\phi_\text{C}\rangle$ transition, which offers a precise measure of trap depth due to its large differential polarizability and narrow linewidth. We ultimately use this signal as feedback to calibrate out imperfections in our imaging onto the CMOS, and to measure uniformity after the iteration is complete.  

Our measured trap depth and radial trap frequency (see Sec.~\ref{section:sideband_cooling}) are consistent with a nearly-diffraction-limited tweezer waist of $\sim$500~nm. We additionally confirm this value by imaging the focal plane of the trap light with an ultra-high resolution objective. However, the observed size of our single-atom point spread function (Fig.~\ref{FigFluorescenceImaging}a, Fig.~\ref{FigTweezerArrays}b) is larger than the theoretical diffraction-limited value.  We suspect thermal spatial broadening, pixelation effects, chromatic shifts between the green trap and blue fluorescence, and/or aberrations in the imaging system to be responsible for this. We leave this for further investigation as this does not directly impact the results presented here.

We observe cooling features across the linearly polarized array similar to those of a single tweezer with magic polarization (Fig.~\ref{FigTweezerArrays}e). We again find a narrow red-detuned cooling feature, but further to the red than that in magic polarization. We expect this feature to be a combination of sideband cooling and Sisyphus cooling in the regime of a more strongly trapped excited state~\cite{Taieb1994, Ivanov2011}. The blue-detuned Sisyphus feature is also still present, albeit extending even further to the blue. These observations are consistent with how we expect excited state polarizabilities to shift with tweezer polarization ellipticity (Fig.~\ref{FigExperimentalSystem}d). For optimal cooling conditions, we again see that the loss coefficient $\chi$ reaches the same minimum value over a broad range of settings (Fig.~\ref{FigTweezerArrays}d), although with a higher value than observed in a single magic tweezer. We leave this observation for further investigation and at this point only hypothesize that it may be partly due to an altered fluorescence radiation pattern because of the difference in tweezer polarization~(App.~\ref{appendix:fluorescence_imaging}).  


\begin{figure}[t!]
	\centering
	\includegraphics[width=\columnwidth]{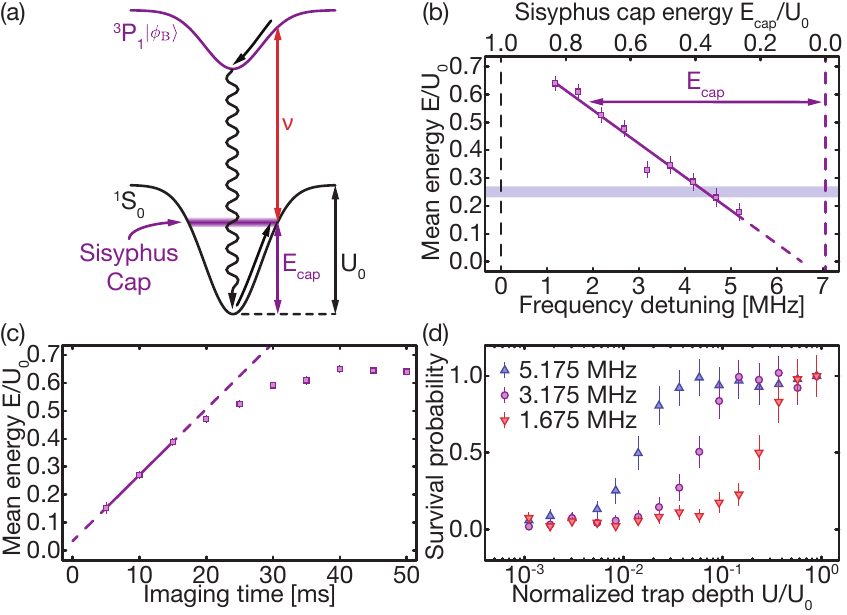}
	\caption{
	\textbf{Sisyphus cooling.}
	{(a)}~Diagram illustrating the mechanism of Sisyphus cooling on the $\ce{^{1}S_{0}}\leftrightarrow\ce{^{3}P_{1}}|\phi_\text{B}\rangle$ transition in the regime where the excited state is less trapped than the ground state. The red cooling beam at frequency $\nu$ is blue-detuned away from free-space resonance, effectively creating a resonance condition for ground state atoms with energy $\text{E}_\text{cap}$. During fluorescence imaging, atoms are heated up until their energy reaches the Sisyphus cap, at which point they are excited and preferentially decay back to the ground state with lower motional energy.
	{(b)}~Mean equilibrium energy of the atom after fluorescence imaging, as a function of the Sisyphus detuning. The solid line is a linear fit to the experimental data. The shaded region represents the equilibrium energy after fluorescence imaging with sideband cooling instead of Sisyphus cooling.
	{(c)}~Mean equilibrium energy of the atom versus imaging time for a Sisyphus detuning of $1.2~\text{MHz}$. The energy initially increases linearly (solid line, $t\leq15~\text{ms}$) and later saturates.  
	{(d)}~Survival probability versus normalized final trap depth after adiabatically ramping down the trap depth, for various Sisyphus cap energies (App.~\ref{appendix:sisyphus_cooling}).
	}
	\label{FigSisyphusCooling}
\end{figure}

\vspace{-3mm}
\section{Sisyphus cooling}~\label{section:sisyphus_cooling}
We now investigate the mechanism behind the broad, blue-detuned cooling feature observed during fluorescence imaging. The feature spans a range of frequencies for which a local resonance condition of the \textit{non-magic} $\ce{^{1}S_{0}} \leftrightarrow \ce{^{3}P_{1}}|\phi_\text{B}\rangle$ transition exists in the trap (Fig.~\ref{FigSisyphusCooling}a). As the red transition is much narrower than the differential trap depth ($\hbar\Gamma \lesssim |\Delta U|$), selective excitation of narrow equipotential manifolds in the trap is possible. By appropriate choice of detuning, an atom can lose energy by exciting on a manifold where the energy of the absorbed photon is smaller than the energy of the photon emitted after oscillating in the excited state potential. This is only effective when the atom spends time in the excited state that is at least commensurate with the trapping period, so the condition $\Gamma \lesssim \omega$ must also hold.  Such a cooling scheme is reminiscent of Sisyphus cooling between ground hyperfine manifolds of alkali atoms~\cite{Adams1997}. Narrow linewidth versions of Sisyphus cooling have been discussed theoretically in Refs.~\cite{Taieb1994, Ivanov2011}, although with the excited state experiencing stronger trapping, which -- as we detail below -- leads to different behavior compared to the case studied here where the excited state experiences weaker trapping.

We measure the equilibrium energy reached during fluorescence imaging with simultaneous Sisyphus cooling and observe a linear dependence on the detuning (Fig.~\ref{FigSisyphusCooling}b). We confirm that an equilibrium is reached by also measuring the mean energy as a function of imaging time and finding that it saturates after an initial linear growth (Fig.~\ref{FigSisyphusCooling}c). These measurements are performed via adiabatic rampdown of the trap to probe the energy distribution~\cite{Tuchendler2008} (Fig.~\ref{FigSisyphusCooling}d and App.~\ref{appendix:sisyphus_cooling}). We quote a mean energy instead of temperature as it is \textit{a priori} not clear whether the reached equilibrium state corresponds to a thermal distribution.

Our interpretation for the linear behavior of mean energy vs detuning is as follows: as atoms scatter blue photons, they heat up, eventually reaching an energy manifold that is resonant with the red cooling light.  Here, Sisyphus cooling counteracts recoil heating.  An equilibrium is reached as recoil heating pushes the energy up against a `Sisyphus cap'. Detunings closer to the free-space resonance, resonant with equipotentials near the top of the trap, result in higher energy caps.  Detunings further to the blue of free-space, resonant with equipotentials deep in the trap, result in lower energy caps.  Consistent with this interpretation, the observed mean energies are slightly below the calculated cap energy, and follow the cap energy in a linear fashion. 

We further observe that if the Sisyphus detuning is suddenly changed to a value further to the blue of what it was upon equilibration of the energy, rapid heating and atomic loss occurs even if blue fluorescence is turned off (not shown).  These observations, which are supported by numerical simulation, paint a broader picture of the Sisyphus mechanism acting as a \textit{repeller} in energy space. That is, atoms with an energy below that of the resonant manifold are pushed to lower energies while atoms with an energy higher than the resonant manifold are heated to even higher energies. We note that we drive a transition such that the excited state experiences weaker trapping than the ground state ($\alpha_e < \alpha_g$). Previous proposals of narrow-line Sisyphus cooling~\cite{Taieb1994, Ivanov2011} have mostly focused on the opposite regime ($\alpha_e > \alpha_g$), in which the Sisyphus mechanism acts as an \textit{attractor} in energy space instead. The latter regime has been proposed as a mechanism for ground state cooling, while our regime is not as amenable to this because cooling stops after the atom has been cooled to some energy that is no longer resonant with the repeller; however, a dynamically swept detuning may achieve very low energies, which we leave for further investigation.

\begin{figure}[t!]
	\centering
	\includegraphics[width=\columnwidth]{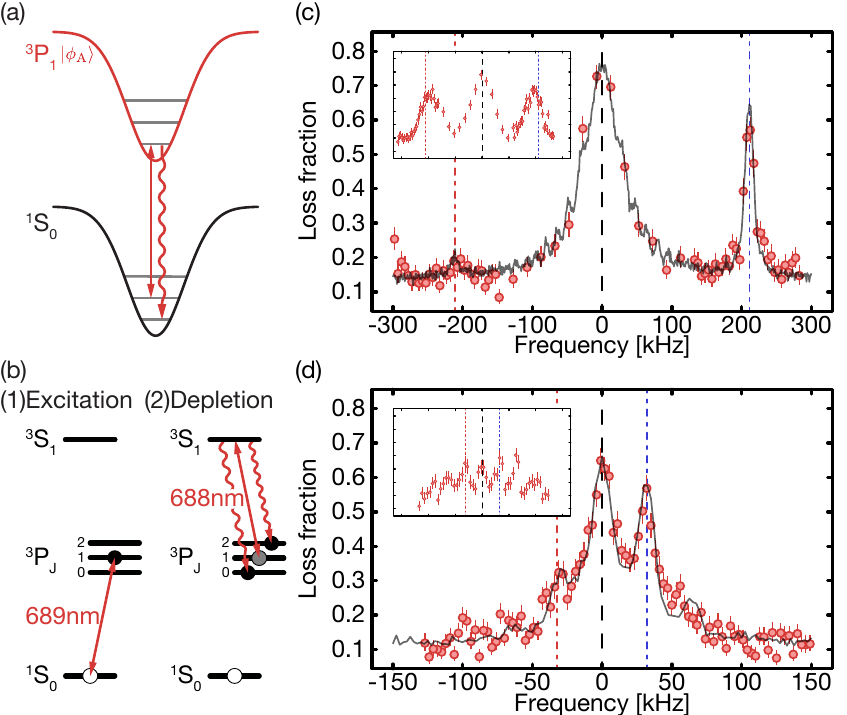}
	\caption{
	\textbf{Sideband cooling.}
	{(a)}~Diagram of the approach to resolved sideband cooling on the magic-tuned $\ce{^{1}S_{0}}\leftrightarrow\ce{^{3}P_{1}}|\phi_\text{A}\rangle$ transition. Optical excitation of the red sideband is spectrally resolved, and the subsequent decay conserves the motional quanta with high probability. 
	{(b)}~Measurement protocol for sideband spectra. Atoms in the $\ce{^{1}S_{0}}$ ground state are excited (solid double arrow) to the $\ce{^{3}P_{1}}$ excited state by an excitation pulse at $689~\text{nm}$. Atoms in the $\ce{^{3}P_{1}}$ excited state are then excited (solid double arrow) to the $\ce{^{3}S_{1}}$ state by a depletion pulse at $688~\text{nm}$, where they radiatively decay to the $\ce{^{3}P_{0}}$ and $\ce{^{3}P_{2}}$ metastable dark states. 
	{(c)}~Radial sideband spectrum before (inset) and after sideband cooling. Overlayed is a simulated spectrum with 0.80 ground state fraction (solid gray line). Bumps visible in the simulated spectrum are Fourier peaks due to the finite $74~\mu\text{s}$ excitation pulse. The first radial sidebands are separated from the carrier frequency by $211(4)~\text{kHz}$. The amplitude of the red sideband is highly suppressed after cooling, as is the width of the blue sideband - both indicating larger ground state fraction. 
	{(d)}~Axial sideband spectrum before (inset) and after the second stage of axial cooling. Overlayed is a simulated spectrum with 0.50 ground state fraction (solid gray line). The first axial sidebands are separated from the carrier frequency by $32.2(8)~\text{kHz}$. Suppression of the red sideband and enhancement of the carrier both indicate larger ground state fraction.  
	}
	\label{FigSidebandCooling}
\end{figure}

\vspace{-3mm}
\section{Sideband cooling in a single tweezer}~\label{section:sideband_cooling}
Finally, we show a proof-of-principle for resolved sideband cooling in a tweezer, hence demonstrating direct optical control of motional degrees of freedom of a tightly trapped single atom. Related work on Raman sideband cooling has been performed with alkali atoms~\cite{Kaufman2012,Thompson2013}, and narrow-line resolved sideband cooling has been previously observed with alkaline-earth(-like) atoms~\cite{Ido2003,Yamamoto2016} and trapped ions~\cite{Diedrich1989,Monroe1995}. Here, we use the $\ce{^{1}S_{0}} \leftrightarrow \ce{^{3}P_{1}}|\phi_\text{A}\rangle$ transition in an elliptically polarized tweezer tuned to the magic angle. The vanishing differential polarizability of this transition simplifies sideband cooling and spectroscopy because sideband transition frequencies do not (up to effects of anharmonicity) depend on the motional state. However, we do not discount the possibility of high-fidelity sideband cooling in finite differential polarizability, and leave this for future studies. 

Since the linewidth of the $\ce{^{1}S_{0}}\leftrightarrow\ce{^{3}P_{1}}$ transition ($7.4~\text{kHz}$) is smaller than our trap frequencies, we can selectively drive red sideband transitions that reduce the motional quantum number (Fig.~\ref{FigSidebandCooling}a). Specifically, for our trap depth of $1.4$~mK (29~MHz), the radial (axial) trap frequency is $\nu_\text{r} = 211(4)~\text{kHz}$ ($\nu_\text{a} = 32.2(8)~\text{kHz}$). Cooling hinges on the propensity for the atom to preserve its motional quantum number while decaying from the excited state, a condition that is achieved when the Lamb-Dicke parameter $\eta$ is small, i.e. $\eta \equiv k\sqrt{\frac{\hbar}{4\pi m\nu}} \ll 1$. For us, the radial direction has $\eta_r = 0.15$ and the axial has $\eta_a = 0.39$.  


Before the start of the cooling sequence, the atom is imaged with Sisyphus cooling and has equilibrated at a mean energy where we expect negligible ground state population (Sec.~\ref{section:sisyphus_cooling}). To cool close to the motional ground state, we perform sideband cooling by alternating 100 $\mu$s pulses of three beams, two orthogonal beams in the radial plane and one beam in the axial direction collimated through our objective. None of the beams are retro-reflected. We break cooling into two stages: the first stage targets the fifth red axial sideband, while the second stage targets the first red axial sideband. Both stages target the first red radial sideband. The first stage is repeated for 100 consecutive cycles, while the second is repeated for 50. 

To extract information about the final motional state, we probe the sideband spectrum after cooling by performing excitation-depletion spectroscopy on the $\ce{^{1}S_{0}}\leftrightarrow\ce{^{3}P_{1}}$ transition~(Fig.~\ref{FigSidebandCooling}b). We first excite the ground state atoms on the $\ce{^{1}S_{0}}\leftrightarrow\ce{^{3}P_{1}}$ transition with an excitation pulse of $74~\mu\text{s}$. We then pump atoms in $\ce{^{3}P_{1}}$ to the $\ce{^{3}P_{0}}$ and $\ce{^{3}P_{2}}$ metastable dark states via the $\ce{^{3}S_{1}}$ state with a depletion pulse of $10~\mu\text{s}$ at $688~\text{nm}$. This excitation-depletion cycle is repeated 3 times to increase signal. Thus, population of $\ce{^{3}P_{1}}$ is measured as apparent loss upon performing a second fluorescence image.  



We observe that a sideband asymmetry appears after cooling (Fig.~\ref{FigSidebandCooling}c,d), which did not exist before cooling (insets), directly demonstrating reduced motional energy. A similar level of asymmetry is observed in the orthogonal radial spectrum (not shown). To quantify the final motional state, we fit our data to simulation of the probe spectroscopy which includes the effect of finite decay (App.~\ref{appendix:sideband_thermometry}). We find our data to be compatible with a thermal ground state fraction in the interval of $[0.69, 0.96]$ in the radial direction and $[0.45, 0.59]$ in the axial. These values refer to the motional state right after sideband cooling, before the probe is applied.

We finally note that we observe a small loss probability during sideband cooling and hypothesize that this may be due to off-resonant excitation from the trapping light while the atom is in $\ce{^{3}P_{1}}$.  Such excitation could induce loss by populating states outside our imaging and cooling cycles.  A longer wavelength trap would likely reduce these losses by being further detuned from higher-lying states.

\vspace{-3mm}
\section{Outlook}\label{sec:outlook}
We have demonstrated trapping, high-fidelity detection, and narrow-line cooling of individual AEAs in optical tweezers. Our imaging technique is based on fluorescence imaging while cooling with a novel narrow-linewidth Sisyphus scheme. 

The robust operation of the Sisyphus mechanism away from finely tuned magic conditions opens the possibility for aiding single-atom imaging in a myriad of situations. Specifically, this presents a viable option for cooling during imaging of essentially any atomic species with sufficiently narrow optical lines, such as other AEAs or dipolar atoms~\cite{Aikawa2012,Lu2011}. As a point of reference, we have demonstrated high-fidelity imaging in trap depths as low as $300~\mu\text{K}$ and anticipate extensions to even shallower depths with further optimization. We note that Sisyphus cooling can be achieved with a single beam as it relies on energy transfer from differential trapping instead of photon momentum. This is often an advantage in such imaging applications as stray light can be minimized.

Concerning strontium itself, Sisyphus cooling can enable imaging in various useful wavelengths. For example, quantum gas microscopes could be operated with $1064~\text{nm}$ light, where high-power lasers exist. Another intriguing possibility is trapping and imaging in $813.4~\text{nm}$, which is a magic wavelength for the $\ce{^{1}S_{0}}\leftrightarrow\ce{^{3}P_{0}}$ clock transition. Importantly, for these wavelengths, we expect that the $\ce{^{1}D_{2}}$ state will be trapped, such that imaging loss from depopulation can be further mitigated by repumping in the triplet and/or singlet manifold.

More broadly, the presented results open the door for a wide range of experimental possibilities enabled by combining OT-based single-atom control techniques with the intriguing features of AEAs. For example, the unique spectral properties of AEAs are currently exploited in optical lattice clocks~\cite{Ludlow2015}. Here, combing single-atom control with such high spectral resolution could be employed to explore systematic shifts introduced by dipole-dipole interactions~\cite{Chang2004} or to implement single-experiment interleaved clock operation~\cite{Schioppo2017}. Further, the combination of long-range interactions mediated by Rydberg states~\cite{Gil2014, Arias2018} or cavity-modes~\cite{Norcia2018} with OTs could be used to controllably introduce and detect entanglement in the clock transition -- a possible pathway to quantum-enhanced clock operation. 

We further note new avenues in quantum simulation and computing. Previously, a combination of high-precision spectral control, unique spin properties~\cite{Pagano2014,Cazalilla2014} and orbital spin exchange interactions~\cite{Scazza2014,Cappellini2014} has been experimentally explored and proposed in a range of AEA quantum simulation applications, including the generation of spin orbit coupling in synthetic dimensions~\cite{Mancini2015,Kolkowitz2017} or work towards Kondo-like systems~\cite{Riegger2018, Gorshkov2010, FossFeig2010}. Related ideas appear in a whole array of quantum computing protocols for AEAs~\cite{Daley2008,Gorshkov2009,Daley2011,Daley2011e}. Specifically, such quantum computing architectures require dedicated single atom control techniques, which could be realized with OTs~\cite{Pagano2018} instead of optical lattices as originally envisioned. In a modification of these protocols, Kondo-type models~\cite{Gorshkov2010, FossFeig2010, FossFeig2010b, Cazalilla2014} could be explored in a bottum-up manner similar to Hubbard models~\cite{Kaufman2014} either with OTs alone or by combining OTs with degenerate quantum gases to introduce impurities.

Further, our experiments will allow control of AEA Rydberg interactions~\cite{Vaillant2012, Mukherjee:2011, Lochead2013, Gil2014, DeSalvo2016, Gaul2016, Bridge2016} at the single atom level, which could lead to an increase in effective coherence time (compared to alkalis) by using meta-stable intermediate states~\cite{Bridge2016, Gil2014} -- an important aspect for further advances in Rydberg-based quantum simulation and computing. 

Finally, we consider OT based strategies for basic atomic physics experiments. For example, we envision controlled ionization of an alkaline-earth atom trapped in a tweezer, providing a new pathway to optical trapping and control of ions~\cite{Schmidt2018}. Further, we note the possibility of generating cold molecules involving AEAs~\cite{Reinaudi2012} in an atom-by-atom fashion using optical tweezers~\cite{Liu2018}. 

\section*{Acknowledgments}\label{sec:acknowledgments}
We acknowledge A. Kaufman and N. Hutzler for insightful discussions. We acknowledge the technical contributions of {Alexander Baumg\"artner} and {Brian Timar}. We acknowledge funding provided by the Institute for Quantum Information and Matter, an NSF Physics Frontiers Center (NSF Grant PHY-1733907). This work was supported by the NSF physics frontier center IQIM, the Sloan Foundation, and by the NASA/JPL President's and Director's Fund. Theoretical work was performed under the sponsorship of the U.S. Department of Commerce, National Institute of Standards and Technology.
\vspace{5mm}
\setcounter{section}{0}
\twocolumngrid

\begin{center}
\textbf{APPENDIX}
\end{center}

\section{Calculation of polarizabilities, magic wavelengths, and branching ratio}~\label{appendix:polarizability}\vspace{-10mm}
\subsection{Overview}
The trapping potential experienced by an atom prepared in its internal state $i$ is equal to the product of the state-dependent polarizability $\alpha_i(\lambda, \hat{\epsilon})$ and the intensity profile of the optical tweezer $I(r,z)$ such that
\begin{eqnarray}
U_i(r,z)&=&-\alpha_i(\lambda, \hat{\epsilon})I(r,z)/2\epsilon_0c,
\end{eqnarray}
where $\epsilon_0$ is the vacuum permittivity and $c$ is the speed of light in vacuum~\cite{Steck2007}. The state-dependent polarizability, $\alpha_i(\lambda, \hat{\epsilon})$, depends on both the wavelength $\lambda$ and the polarization vector $\hat{\epsilon}$ of the trapping light~\cite{Ido2003, Steck2007}. The polarizability of the $\ce{^{1}S_{0}}$ ground state is independent of polarization, whereas the polarizabilities of the three sublevels of the $\ce{^{3}P_{1}}$ excited state depend on the polarization due to vector and tensor components of the polarizability.

\begin{figure}[t!]
\includegraphics[width=\columnwidth]{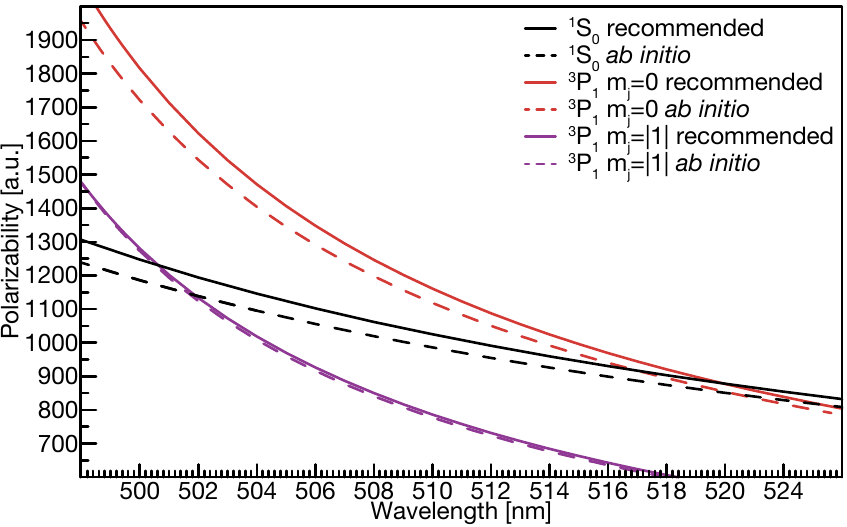}
\caption{
\textbf{Polarizabilities of the $5\text{s}^2~\ce{^{1}S_{0}}$ and $5\text{s}5\text{p}~\ce{^{3}P_{1}}$ states for $\ce{Sr}$ at linear trap polarization.}
Calculations with both \textit{ab initio} (dashed lines) and recommended (solid lines) values predict the same magic wavelength at $520(2)~\text{nm}$ for the
$\ce{^{1}S_{0}}\leftrightarrow\ce{^{3}P_{1}}|m_j=0\rangle$ transition with $ \alpha=880(25)~\text{a.u.}$ Calculations with recommended values predict another magic wavelength at $\lambda=500.65(50)~\text{nm}$ for the $\ce{^{1}S_{0}}\leftrightarrow\ce{^{3}P_{1}}|m_j=\pm1\rangle$ transition with $\alpha=1230(13)~\text{a.u.}$ We note that this latter crossing is valid even for elliptical trap polarizations, as it pertains to an excited sublevel with polarization-insensitive polarizability.  
}
\label{sFigMagicWavelength}
\end{figure}

We calculate the polarizability of the $\ce{^{1}S_{0}}$ and $\ce{^{3}P_{1}}$ states (Fig.~\ref{sFigMagicWavelength}) using both \emph{ab initio} and recommended values for the transition wavelengths and dipole matrix elements~(see Table~\ref{tab1} for the computed and recommended values, as well as the breakdown of contributions to the polarizability). At linear trap polarization, we predict a magic wavelength on the $\ce{^{1}S_{0}}\leftrightarrow\ce{^{3}P_{1}}|m_j^x=0\rangle$ transition at $520(2)~\text{nm}$ using both \emph{ab initio} and recommended values. We predict another magic wavelength on the $\ce{^{1}S_{0}}\leftrightarrow\ce{^{3}P_{1}}|m_j^x=\pm1\rangle$ transition at $\lambda=500.65(50)~\text{nm}$ using recommended values. 

The wavelength of our tweezers is $515.2~\text{nm}$, such that for linear polarization the trapping potential in the $\ce{^{3}P_{1}}|m_j^x=0\rangle$ ($|m_{j}^{x}=\pm{1}\rangle$) excited state is larger (smaller) than the trapping potential in the $\ce{^{1}S_{0}}$ ground state by $5~\%$ ($30~\%$). We achieve a magic trapping condition by tuning to elliptical polarization as detailed in App.~\ref{appendix:tuning}. 
\vspace{-3mm}
\subsection{Calculating polarizabilities and magic wavelengths for Sr}
The frequency-dependent scalar polarizability, $\alpha(\omega)$, of an
atom in a state $i$ may be separated into a core polarizability
 $\alpha_{\rm{core}}$ and a contribution from the valence electrons,
$\alpha^v(\omega)$. The core polarizability is a sum of the polarizability of the ionic $\ce{Sr^{2+}}$ core and a counter term that
compensates for Pauli principle violating core-valence excitation from the core to the valence shells. 
The ionic core polarizability is small and a static value calculated in the random-phase approximation (RPA) gives sufficient accuracy \cite{Safronova2013}.

The total polarizability for linear polarization is given by
\begin{eqnarray}
\alpha=\alpha_s+\alpha_t \frac{3m_j^2-J_i(J_i+1)}{J_i(2J_i-1)},
\end{eqnarray}
where $J_i$ is the total angular momentum quantum number of the state $i$, $m_j$ is the magnetic quantum number associated with the projection of the angular momentum along the polarization axis of the tweezer ($\hat{x}$), and $\alpha_s$ and $\alpha_t$ are the scalar and tensor polarizabilities, respectively.
 The total polarizability for the $J_i=1$ state is
given by
\begin{eqnarray}
\alpha=\alpha_s-2\alpha_t
\end{eqnarray}
for $m_j=0$
and
\begin{eqnarray}
\alpha=\alpha_s+\alpha_t
\end{eqnarray}
for $m_j=\pm 1$.

We calculate the valence polarizabilities using a hybrid approach that combines configuration iteration
and a linearized coupled-cluster method [CI+all-order] \cite{Safronova2009}. The application of this method to the calculation of polarizabilities is
described in Refs.~\cite{Safronova2013,Porsev2014}. Briefly,
the valence part of the polarizability for the state $i$ with the total angular momentum $J_i$ and projection
$m_j$ is determined by solving the inhomogeneous equation of perturbation theory in the
valence space, which is approximated as \cite{Porsev1999}
\begin{equation}
(E_v - \mathcal{H}_{\textrm{eff}})|\Psi(v,m_j^{\prime})\rangle = D_{\mathrm{eff},q} |\Psi_0(v,J_i,m_j)\rangle.
\label{eq1}
\end{equation}
The parts of the wave function $\Psi(v,m_j^{\prime})$
with angular momenta of $J_i^{\prime} = J_i, J_i \pm 1$ are then used  to determine
the scalar and tensor polarizabilities.  The $\mathcal{H}_{\textrm{eff}}$ includes the all-order corrections calculated using the
linearized coupled-cluster method with single and double excitations.
The effective dipole operator $D_{\textrm{eff}}$ includes RPA corrections.
This approach automatically includes contributions from all possible states.
 \begin{table} \caption[]{Contibutions to the $\ce{Sr}$ scalar $\alpha_s$ and tensor $\alpha_t$ polarizabilities for the $5\text{s}^2~\ce{^{1}S_{0}}$ and $5\text{s}5\text{p}~\ce{^{3}P_{1}}$ states at 520~nm and 515.2~nm in a.u. Correspoding energy differences $\Delta E$ in cm$^{-1}$ and reduced electric-dipole matrix elements $D$ in a.u. are also listed.}\label{tab1}
\begin{ruledtabular}
\begin{tabular}{lcrrrrr}
   \multicolumn{1}{c}{Contribution} &
    \multicolumn{2}{c}{} &
      \multicolumn{2}{c}{520~nm} &
          \multicolumn{2}{c}{515.2~nm} \\
\multicolumn{1}{c}{} &
\multicolumn{1}{c}{$\Delta E$} &
  \multicolumn{1}{c}{$D$} &
    \multicolumn{1}{c}{$\alpha_s$} &
      \multicolumn{1}{c}{$\alpha_t$} &
          \multicolumn{1}{c}{$\alpha_s$} &
      \multicolumn{1}{c}{$\alpha_t$} \\
[0.3pc] \hline
  \multicolumn{7}{l}{$5\text{s}^2~\ce{^{1}S_{0}}$ polarizability}  \\   \hline
 $5\text{s}5\text{p}~\ce{^{3}P_{1}}$ &14504    &0.151   &   -0.3&       & -0.3  &    \\
 $5\text{s}5\text{p}~\ce{^{1}P_{1}}$ &21698    &5.248(2)&  865.7&       & 929.4 &     \\
             Other         &         &        & 7.2   &       &  7.3  &    \\
             Core          &         &        & 5.3   &       &   5.3 &    \\
             \textbf{Total}         &         &        & \textbf{878.0} &       &\textbf{941.8}  &     \\ [0.3pc]
            \hline
             \multicolumn{7}{l}{$5\text{s}5\text{p}~\ce{^{3}P_{1}}$ polarizability}     \\           \hline
$5\text{s}^2~\ce{^{1}S_{0}}$&    -14504 &   0.151    &  0.1   &  -0.1  &  0.1   & -0.1       \\
$5\text{s}4\text{d}~\ce{^{3}D_{1}}$&    3655   & 2.322(11)  &  -2.7  &  -1.3  &  -2.6  &  -1.3     \\
$5\text{s}4\text{d}~\ce{^{3}D_{2}}$&    3714   & 4.019(20)  &  -8.2  &   0.8  &  -8.1  &  0.8       \\
$5\text{s}4\text{d}~\ce{^{1}D_{2}}$&    5645   & 0.190      &  0.0   &   0.0  &  0.0   &  0.0       \\
$5\text{s}6\text{s}~\ce{^{3}S_{1}}$&    14534  &  3.425(17) &   -52.4&  -26.2 & -50.2  &  -25.1     \\
$5\text{s}6\text{s}~\ce{^{1}S_{0}}$&    16087  &  0.045     &   0.0  &   0.0  &  0.0   &  0.0        \\
$5\text{s}5\text{d}~\ce{^{1}D_{2}}$&    20223  &  0.061     &    0.1 &   0.0  &  0.1   &  0.0       \\
$5\text{s}5\text{d}~\ce{^{3}D_{1}}$&    20503  &  2.009(20) &   79.9 &   39.9 &   92.5 &   46.3      \\
$5\text{s}5\text{d}~\ce{^{3}D_{2}}$&    20518  &  3.673(37) &   263.9& -26.4  &  305.2 &   -30.5     \\
$5\text{p}^2~\ce{^{3}P_{0}}$&    20689  &  2.657(27) &   122.4& -122.4 &   138.9&    -138.9   \\
$5\text{p}^2~\ce{^{3}P_{1}}$&    20896  &  2.362(24) &   85.1 &   42.6 &   94.9 &   47.5       \\
$5\text{p}^2~\ce{^{3}P_{2}}$&    21170  &  2.865(29) &   108.2&  -10.8 &   118.6&    -11.9    \\
$5\text{p}^2~\ce{^{1}D_{2}}$&    22457  &  0.228     &    0.4 &   0.0  &  0.4   &  0.0         \\
$5\text{p}^2~\ce{^{1}S_{0}}$&    22656  &  0.291     &   0.7  &   -0.7 &   0.7  &  -0.7        \\
$5\text{s}7\text{s}~\ce{^{3}S_{1}}$&    22920  &  0.921     &   6.1  &    3.0 &   6.4  &  3.2         \\
              Other        &           &            &  65.8  &    0.2 &   66.9 &   0.2         \\
              Core         &           &            &  5.6   &    0.0 &   5.6  &  0.0         \\
              \textbf{Total}        &           &            &  \textbf{674.7} &  \textbf{-101.3}&  \textbf{769.4} &   \textbf{-110.5}      \\
   \end{tabular}
   \end{ruledtabular}
   \end{table}
To improve accuracy, we extract several contributions
to the valence polarizabilities using the sum-over-states formulas \cite{Mitroy2010}:
\begin{eqnarray}
    \alpha_{s}(\omega)&=&\frac{2}{3(2J_i+1)}\sum_k\frac{{\left\langle k\left\|D\right\|i\right\rangle}^2(E_k-E_i)}{     (E_k-E_i)^2-\omega^2}, \label{eq-1} \nonumber \\
    \alpha_{t}(\omega)&=&4C\sum_k(-1)^{J_i+J_k}
            \left\{
                    \begin{array}{ccc}
                    J_i & 1 & J_k \\
                    1 & J_i & 2 \\
                    \end{array}
            \right\} \nonumber \\
      & &\times \frac{{\left\langle
            k\left\|D\right\|i\right\rangle}^2(E_k-E_i)}{
            (E_k-E_i)^2-\omega^2} \label{eq-pol},
\end{eqnarray}
             where $C$ is given by
\begin{equation}
C =\left(\frac{5J_i(2J_i-1)}{6(J_i+1)(2J_i+1)(2J_i+3)}\right)^{1/2} \nonumber
\end{equation}
We calculate two such contributions for the $\ce{^{1}S_{0}}$ polarizability and 15 contributions for the $\ce{^{3}P_{1}}$ polarizability with \textit{ab initio} energies and matrix elements that exactly correspond to our calculations using the inhomogeneous Eq.~(\ref{eq1}) and determine the remainder contribution of all other states.  Then we do the same calculation using the experimental energies and recommended values of matrix elements from Ref.~\cite{Porsev2014} where available.
The recommended value for the $\ce{^{1}S_{0}}\leftrightarrow\ce{^{1}P_{1}}$ matrix element is from the $\ce{^{1}P_{1}}$ lifetime measurement~\cite{Yasuda2006}. We add the core and the remainder contribution from the other states (labeled as ``Other'' in Table~\ref{tab1}) to these values to obtain the final results.

The results of this calculation for 520~nm and 515.2~nm are listed in Table~\ref{tab1} in atomic units (a.u.), as well as the energy difference $\Delta E=E_k-E_i$ in cm$^{-1}$ and the absolute values of the reduced electric-dipole matrix elements $D$ in $a_0 |e|$ (a.u.), where $a_0$ is the Bohr radius and $e$ is the elementary charge. The core and remainder contributions are also listed\footnote{
We use the conventional system of atomic units,
a.u., in which $e$, the electron mass $m_{\rm e}$, and the reduced
Planck constant $\hbar$ have the numerical value 1, and the electric constant $\epsilon_0$ has the numerical value $1/(4\pi)$. 
The atomic units for $\alpha$ can be converted to SI units via $\alpha/h$~[Hz/(V/m)$^2$]=2.48832$\times10^{-8}\alpha$~[a.u.], where the conversion coefficient is $4\pi \epsilon_0 a^3_0/h$ and the Planck constant $h$ is factored out.}.
 We carry out the same calculations for the other wavelengths to determine the magic wavelengths for which $\ce{^{1}S_{0}}$ and $\ce{^{3}P_{1}}$ polarizabilities have the same values. The results of the \textit{ab initio} calculation and the calculations corresponding to Table~\ref{tab1} (recommended) are illustrated in Fig.~\ref{sFigMagicWavelength}.

\vspace{-3mm}
\subsection{Calculating the Q value}
\begin{table} \caption[]{Polarizabilities and $Q$ values of the $5\text{s}^2~\ce{^{1}S_{0}}$ and $5\text{s}5\text{p}~\ce{^{3}P_{1}}$ states in a.u. at 515.2~nm for Sr. The recommended $Q$ values are obtained using the polarizability values provided in Table~\ref{tab1}. The $Q$ values listed in the row labeled ``Expt. energy'' are obtained using the experimental energies and theoretical matrix elements.}
\label{tab2}
\begin{ruledtabular}
\begin{tabular}{lcccccc}
   \multicolumn{1}{l}{Method} &
    \multicolumn{1}{c}{$\alpha(^1S_0)$} &
   \multicolumn{1}{c}{$\alpha_s(^3P_1)$} &
      \multicolumn{1}{c}{$\alpha_t(^3P_1)$} &
\multicolumn{2}{c}{$\alpha(^3P_1)$} &
  \multicolumn{1}{c}{$Q$} \\
\multicolumn{4}{c}{} &
\multicolumn{1}{c}{$m_j=0$} &
  \multicolumn{1}{c}{$m_j=\pm1$} &
  \multicolumn{1}{c}{} \\
  \hline    
\textit{Ab initio}&	910	&754	&-103	&960	&651	&-5.1\\
Expt. energy    	&951&	776&	-113&	1002	&664	&-5.6\\
Recm.	          & 942	&769	&-111	&990	&659&	-5.8\\
 \end{tabular}
   \end{ruledtabular}
   \end{table}

We use the polarizability results to calculate the $Q$ value, defined as the ratio of differential polarizabilities
\begin{equation}~\label{eq:Q_value}
Q=\frac{\alpha(\ce{^{1}S_{0}}) - \alpha(\ce{^{3}P_{1}}|m_j=\pm1\rangle)}{\alpha(^1S_0) - \alpha(\ce{^{3}P_{1}}|m_j=0\rangle)}.
\end{equation} 
Our results are summarized in Table~\ref{tab2}. We note that varying the recommended matrix elements $D$ within their estimated uncertainties $\Delta D$, i.e., using the $D+\Delta D$ and $D-\Delta D$  values of the matrix elements, gives $Q=-4$ and $Q=-10$ values despite only 
$2~\%$ changes in the $\ce{^{3}P_{1}}$ polarizabilities. 
Therefore, $Q$ is an excellent new benchmark of the theoretical methodologies, since it is extremely sensitive to even small changes in several matrix elements. We note that only the uncertainties in the values of 5 matrix elements, $5\text{s}5\text{d}~\ce{^{3}D_{1,2}}$ and $5\text{p}^2~\ce{^{3}P_{0,1,2}}$, contribute significantly to the uncertainty of the $Q$ value. 
We compare the theoretical Q-value to experimental measurements in App.~\ref{appendix:tuning}C.

\vspace{-3mm}
\subsection{Calculating the branching ratio}
We obtain $\langle\ce{^{1}D_{2}}||D||\ce{^{1}P_{1}}\rangle  = 1.956$~a.u. in the CI+all-order approximation 
with RPA corrections to the effective dipole operator. Including other small corrections described in Ref.~\cite{Safronova2013}
yields the final value 
$\langle\ce{^{1}D_{2}}||D||\ce{^{1}P_{1}}\rangle= 1.92(4)$~a.u.
The E1 transition rate $A$ is determined using
\begin{equation}
 A = \frac{2.02613\times 10^{18}} {(2J_a+1)\lambda ^{3}} \, \, S(E1),\\
 \end{equation}
where the transition wavelength $\lambda$ is in \AA~and the line strength $S$ is in atomic units.
Using 
$\langle ^1S_0 \|D\| ^1P_1\rangle  = 5.248(2)$~a.u. we obtain 
\begin{eqnarray}
A(\ce{^{1}P_{1}}\rightarrow\ce{^{1}D_{2}})&=& 9.25(40) \times 10^3~\text{s}^{-1}\\
A(\ce{^{1}P_{1}}\rightarrow\ce{^{1}S_{0}})&=& 1.9003(15) \times 10^8~\text{s}^{-1}.
\end{eqnarray}
The resulting ratio is
\begin{equation}
\frac{A(\ce{^{1}P_{1}}\rightarrow\ce{^{1}S_{0}})}{A(\ce{^{1}P_{1}}\rightarrow\ce{^{1}D_{2}})}\approx20500(900).
\end{equation}
\section{Experimental tuning and measurement of polarizabilities}~\label{appendix:tuning}
\vspace{-10mm}
\subsection{Polarizability tuning with elliptical polarization}
The dependence of polarizability (and hence trap depth) on trap polarization can be derived analytically by solving for the eigenvalues of the AC Stark Hamiltonian~\cite{Steck2007, Rosenband2018}. We being by writing the optical trapping field in a particular point in space as 
\begin{eqnarray}
{\vec{E}(t)}=\vec{E}^{(+)}e^{-i\omega t}+\vec{E}^{(-)}e^{+i\omega t},
\end{eqnarray}
where $\vec{E}^{(+)}=E_0\hat{\epsilon}$, $\vec{E}^{(-)}$ is the complex conjugate of $\vec{E}^{(+)}$, and $\hat{\epsilon}$ is the complex unit polarization vector. We parametrize the ellipticity of $\hat{\epsilon}$ by the ellipticity angle $\gamma$ ~\cite{ Rosenband2018}, writing
\begin{eqnarray}
{\hat{\epsilon}}(\gamma)=\cos{(\gamma)}~\hat{x}+i\sin{(\gamma)}~\hat{y}.
~\label{Eq:PolarizationEllipticity}
\end{eqnarray}
Here, we use a Cartesian coordinate system defined by the unit vectors $\{\hat{x}, \hat{y}, \hat{z}\}$, with $\hat{z}$ oriented along the $\vec{k}$ vector of the optical tweezer. We neglect axial components and spatial variation of the polarization caused by non-paraxial effects near the focal plane~\cite{Thompson2013}. Linear polarization is given by $\gamma=0$ and circular by $\gamma=\pi/4$. 

The trapping field acts as a perturbation to the bare atomic Hamiltonian, causing energy shifts (often referred to as AC Stark shifts or light shifts) and mixing of electronic levels.  Using second-order time-dependent perturbation theory, and after organizing terms into a scalar, vector, and tensor contribution, we can write the perturbation on a particular sublevel manifold as a time-independent AC Stark Hamiltonian~\cite{Steck2007}: 
\begin{eqnarray}
&\mathcal{H}&=
-\alpha_s~ E_0^2~\mathds{1}
+\mu_\text{B}g_{J}\vec{B}_\text{eff}(\alpha_{v})\cdot\vec{J}~\label{Eq:StarkHamiltonian}\\
&-&\frac{3\alpha_{t}}{J(2J-1)}\bigg(\frac{1}{2}\{\vec{E}^{(+)}\cdot\vec{J},\vec{E}^{(-)}\cdot\vec{J}\}
-\frac{1}{3}J(J+1)E_0^2\bigg)\nonumber,
\end{eqnarray}
where $\{\cdot,\cdot\}$ is the anticommutator, $\alpha_{s}$, $\alpha_{v}$, and $\alpha_{t}$ are the scalar, vector, and tensor polarizabilities, $g_J$ is the Land\'e $g$-factor, $\vec{B}_\text{eff}$ is an effective magnetic field (discussed below), and $\vec{J}$ is a vector whose components are the angular momentum operators. Here, we constrain ourselves to the $\ce{^{3}P_{1}}$ sublevel manifold which has $J = 1$.  Hence, in our case $\mathcal{H}$ is a 3$\times$3 matrix.  

We define the effective magnetic field in Eq.~(\ref{Eq:StarkHamiltonian}) as
\begin{eqnarray}
\vec{B}_\text{eff}(\alpha_{v})&=&-\frac{\alpha_{v}}{\mu_\text{B}g_J J}~i(\vec{E}^{(-)}\times\vec{E}^{(+)})\\
~&=&\frac{\alpha_{v}E_0^2}{\mu_\text{B}g_J J}~\sin{(2\gamma)}~\vec{e}_z~\label{Eq:EffectiveField}.
\end{eqnarray}
This term, which is nonzero when the polarization has any ellipticity, induces a perturbation identical to that of a magnetic field perpendicular to the plane of ellipticity (in our case, along $\hat{z}$). Writing the Stark Hamiltonian in this way makes it easy to add the contribution of some external real magnetic field $\vec{B}_0$ by replacing $\vec{B}_\text{eff}(\alpha_v)$ with $\vec{B}_\text{tot}=\vec{B}_\text{eff}(\alpha_v)+\vec{B}_0$. 

In the absence of external magnetic field ($\vec{B}_0=0$), the eigenvalues of the Stark Hamiltonian are given by
\begin{eqnarray}
h\nu_\text{C}(\gamma)&=&-(\alpha_{s}+\alpha_{t})\cdot E_0^2~\label{eq:stark_frequency_c}\\
h\nu_\text{B}(\gamma)&=&-(\alpha_{s}-\left (\alpha_{t}-f(\alpha_v, \alpha_t; \gamma)\right)/2)\cdot E_0^2~\label{eq:stark_frequency_b}\\
h\nu_\text{A}(\gamma)&=&-(\alpha_{s}-\left (\alpha_{t}+f(\alpha_v, \alpha_t; \gamma)\right)/2)\cdot E_0^2~\label{eq:stark_frequency_a},
\end{eqnarray}
where
\begin{eqnarray}
f(\alpha_v, \alpha_t; \gamma)&=&\sqrt{9\alpha_{t}^2\cos ^2(2 \gamma)+4\alpha_{v}^2\sin ^2(2 \gamma )},
\end{eqnarray}
is a mixing factor that depends on the vector polarizability, tensor polarizability, and ellipticity angle. Analytical formulas for the corresponding eigenvectors are possible for a quantization axis along $\hat{z}$, and are given, in unnormalized form, by
\begin{align}
&|\phi_\text{C}(\gamma)\rangle &=& &~&|m_j^z=0\rangle \\
&|\phi_\text{B}(\gamma)\rangle &=&  &~&g_{-}(\gamma)|m_j^z=+1\rangle+|m_j^z=-1\rangle \\
&|\phi_\text{A}(\gamma)\rangle &=& &-&g_{+}(\gamma)|m_j^z=+1\rangle+|m_j^z=-1\rangle,
\end{align}
where 
\begin{eqnarray}
g_{\pm}(\gamma)&=&\frac{f(\alpha_v,\alpha_t;\gamma) \pm 2\alpha_v\sin{(2\gamma)}}{3\alpha_t\cos{(2\gamma)}}.
\end{eqnarray}
The $|\phi_\text{C}(\gamma)\rangle=|m_j^z=0\rangle$ eigenstate is independent of the ellipticity angle as is its corresponding eigenvalue, whereas the $|\phi_\text{B}(\gamma)\rangle$ and $|\phi_\text{A}(\gamma)\rangle$ eigenstates depend on the polarization ellipticity due to mixing of the bare $|m_j^z=\pm1\rangle$ sublevels by the optical field.

For the special case of linear polarization ($\gamma=0$), we have $f(\alpha_v, \alpha_t; 0)=3\alpha_t$ and $g_{\pm}(0)=1$, such that the eigenvalues are given by
\begin{eqnarray}
h\nu_\text{C}(0)&=&-(\alpha_{s}+\alpha_{t})\cdot E_0^2\\
h\nu_\text{B}(0)&=&-(\alpha_{s}+\alpha_{t})\cdot E_0^2\\
h\nu_\text{A}(0)&=&-(\alpha_{s}-2\alpha_{t})\cdot E_0^2.
\end{eqnarray}
The unnormalized eigenvectors for a quantization axis chosen along the propagation axis of the tweezer ($\hat{z}$) are given by
\begin{align}
&|\phi_\text{C}(0)\rangle=~|m_j^z=0\rangle \\
&|\phi_\text{B}(0)\rangle=~|m_j^z=+1\rangle - |m_j^z=-1\rangle\\
&|\phi_\text{A}(0)\rangle=-|m_j^z=+1\rangle + |m_j^z=-1\rangle.
\end{align}
A more common choice of quantization axis (used in App.~\ref{appendix:polarizability}) is along the tweezer polarization ($\hat{x}$).  For this choice, it is also more convenient to choose a different basis in the subspace of the degenerate $|\phi_\text{B}\rangle$ and $|\phi_\text{C}\rangle$ states, such that we can equivalently write (up to degeneracy)
\begin{align}
&|\phi_\text{C}(0)\rangle= |m_j^x=\pm1\rangle\\
&|\phi_\text{B}(0)\rangle= |m_j^x=\mp1\rangle\\
&|\phi_\text{A}(0)\rangle= |m_j^x=0\rangle.  
\end{align}

In the presence of an external longitudinal magnetic field, $\vec{B}_0=B_z\hat{z}$, an analytical form for the eigenvalues and eigenvectors of the Stark Hamiltonian can be obtained by replacing the vector polarizability by
\begin{eqnarray}
\alpha_v \rightarrow (\alpha_v + \frac{g_J\mu_\text{B}B_z}{E_0^2\sin{(2\gamma)}}).
\end{eqnarray}
This would be observed as an asymmetry in the energy spectra between left and right-handed ellipticities.  We measure this asymmetry in our spectra and find it to be consistent with a longitudinal magnetic field on the order of $\sim$15~mG.  It is also possible to diagonalize the Stark Hamiltonian in the presence of transverse magnetic fields (i.e. in $\hat{x}$ or $\hat{y}$), although the resulting formulas are cumbersome. A transverse field would cause splitting of the otherwise degenerate $|\phi_\text{B}\rangle$ and $|\phi_\text{C}\rangle$ eigenstates at linear polarization ($\gamma = 0$). Within our precision, we do not observe such a splitting and conclude that external transverse fields are sufficiently well-nullified.  

\begin{figure}[t!]
	\includegraphics[width=\columnwidth]{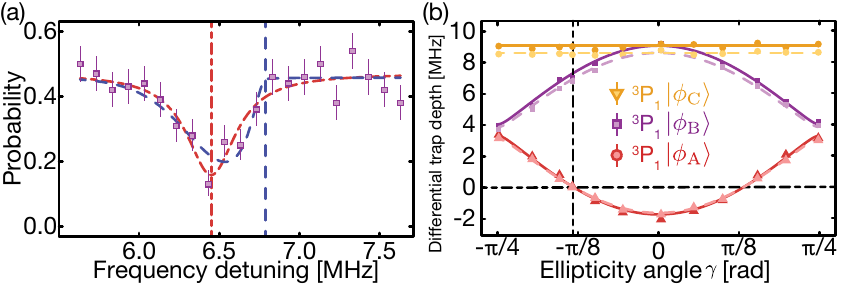}
	\caption{
	\textbf{Differential trap depth spectroscopy.}
	{(a)}~Spectroscopy signal measured on the $\ce{^{1}S_{0}}\leftrightarrow\ce{^{3}P_{1}}|\phi_B\rangle$ transition for the non-magic ellipticity angle $\gamma=28^\circ$. The signal is fitted to a purely thermally broadened line shape (blue dashed curve) and to a purely power broadened line shape (red dash-dot curve). The vertical lines indicate the position of the edge frequency (blue dashed line) and center frequency (red dash-dot line). We expect the true rescaled differential trap depth to lie between these two values as a combination of power and thermal broadening determines the true lineshape.
	{(b)}~Differential trap depth measured on the three $\ce{^{1}S_{0}}\leftrightarrow\ce{^{3}P_{1}}$ transitions for various ellipticity angles assuming a thermally broadened line shape (dark markers) and a power broadened line shape (light markers). The measured values are simultaneously fitted to the analytical solution of the eigenvalues of the Stark Hamiltonian with three free parameters. At the magic ellipticity angle $|\gamma|=24^\circ$ (black dash-dot line), the differential trap depth on the $\ce{^{1}S_{0}}\leftrightarrow\ce{^{3}P_{1}}|\phi_A\rangle$ transition vanishes.
	}
	\label{sFigLightShift}
\end{figure}

\subsection{Measuring the differential trap depth}

We measure the differential trap depth as a function of the ellipticity $\gamma$ by performing excitation-depletion spectroscopy~(Fig.~\ref{FigSidebandCooling}b) on the $\ce{^{1}S_{0}}\leftrightarrow\ce{^{3}P_{1}}$ transitions and fitting the spectroscopy signal to a thermally broadened and power broadened spectral line~(Fig.~\ref{sFigLightShift}a). Specifically, we assume the spectroscopy signal measured after $n$ repetitions of the excitation-depletion cycle to be expressed by $S_n(\nu)=S_0\cdot(1-p_{t}(\nu))^n$, where $S_0$ is the baseline signal measured in the absence of excitation-depletion pulses and $p_{t}(\nu)$ is the probability of pumping the atom from the ground state into a metastable dark state following a single excitation-depletion cycle. We further assume the transition probability to be proportional to the thermal energy distribution in the $\ce{^{1}S_{0}}$ ground state, i.e., $p_{t}(\nu)\propto f(E(\nu))\Theta(E(\nu))$, where $f(E)=\frac{1}{2}\frac{1}{(k_\text{B}T)^3}E^2 e^{-E/{k_\text{B}T}}$ is the Boltzmann energy distribution for a three-dimensional harmonic oscillator and $\Theta(E)$ is the Heaviside function, which restricts the evaluation of the function to positive energy values. 

The resonance condition for an atom at energy $E$ can be written as $E (1-\frac{\alpha_e}{\alpha_g})=\Delta U - h \Delta \nu $. Here, $\alpha_e$ and $\alpha_g$ are the polarizabilities of the excited and ground state, respectively. The differential trap depth is $\Delta U$ and the detuning from free space resonance is $\Delta \nu$. Importantly, when the detuning matches the differential trap depth, $E$ is zero. Hence the edge of the thermal distribution yields the differential trap depth.
Using this approach, we fit the spectroscopy signal to the thermally broadened spectral line and extract the differential trap depth (Fig.~\ref{sFigLightShift}). To account for possible estimation errors associated with power broadening, we further fit the spectroscopy signal to a purely power broadened spectral line, $\tilde{S}_n(\nu)\propto g(\nu)$, where $g(\nu)$ is a normalized Lorentzian function. The mean of the Lorentzian fit provides a bound on the differential trap depth extracted from the cut-off edge that we use as a systematic error bar in Fig.~\ref{FigExperimentalSystem}d. (Even in the limit of extreme power broadening we expect the true value between the edge frequency and the center frequency of the Lorentzian fit.) Were the saturation parameter precisely known from independent measurements, the signal could be fit to a composite lineshape using $\tilde{S}_n(\nu)\propto S_n(\nu)\ast g(\nu)$.

\subsection{Comparing polarizabilities between measured and computed values}
We use the analytical form of the light shifts from Eq.~(\ref{eq:stark_frequency_c}-\ref{eq:stark_frequency_a}) to simultaneously fit our experimental measurements of the differential trap depth~(Fig.~\ref{sFigLightShift}b) using the three free parameters $\{\alpha_s,\alpha_v, \alpha_t\}$. Without any assumptions on $E_0^2$ or $\alpha_g$, we can estimate the $Q$ value defined in Eq.~(\ref{eq:Q_value}) from
\begin{eqnarray}
Q&=&\frac{\Delta \nu_\text{C}(0)}{\Delta \nu_\text{A}(0)}\\
~&=&\frac{(\alpha_g-\alpha_C(0))E_0^2}{(\alpha_g-\alpha_A(0))E_0^2}\\
~&=&\frac{\alpha(\ce{^{1}S_{0}}) - \alpha(\ce{^{3}P_{1}}|m_j^x=\pm1\rangle)}{\alpha(^1S_0) - \alpha(\ce{^{3}P_{1}}|m_j^x=0\rangle)},
\end{eqnarray}
where $\alpha_C(0)=\alpha_s+\alpha_t$ and $\alpha_A(0)=\alpha_s - 2\alpha_t$. The measured $Q=-5.1(3)$ value is consistent with the $Q\in[-5.8,-5.1]$ values estimated from our calculation of the polarizabilities (Table~\ref{tab2}).

In addition, without any assumptions on $E_0^2$ or $\alpha_g$, we can extract the quantity $\abs{\alpha_v}/\abs{\alpha_t}=0.10(4)$ from $\Delta\nu_{BA}(\gamma)/\Delta\nu_{BA}(0)$, where $\Delta \nu_{BA}(\gamma)=\Delta \nu_{B}(\gamma)-\Delta \nu_{A}(\gamma)=f(\abs{\alpha_v},\abs{\alpha_t};\gamma)E_0^2/h$. 



\section{Experimental system}~\label{appendix:experimental_system}
Our scientific apparatus has two ultra-high vacuum regions: the first region is a high flux atomic beam oven and Zeeman slower for strontium (AOSense, Inc.) with integrated transverse cooling in a two-dimensional magneto-optical trap (MOT); the second region is a large stainless steel chamber connected to a glass cell (Japan Cell) in which experiments are carried out. We observe vacuum lifetimes of up to $60~\text{s}$ in a magnetic trap loaded by optically pumping atoms to the metastable $\ce{^{3}P_{2}}$ state.  

We utilize four laser systems: a blue laser system, a red laser system, a repumping laser system, and a green trapping laser system. The blue laser system (Toptica Photonics, TA-SHG Pro System) is a 922~nm diode laser amplified by a tapered amplifier (TA) and frequency doubled in a bow-tie second harmonic generation (SHG) cavity. The red laser system is a 689~nm diode laser (Toptica Photonics, DL pro) locked to a high finesse optical cavity (Stable Laser Systems) and amplified with a home-built TA with a maximum output power of $500~\text{mW}$. The green trapping laser system has a 10~W fiber laser (Azur Light Systems) at $515.2~\text{nm}$ operated in free space without any additional fibers. The repumping laser system has three diode lasers stabilized by a wavemeter (HighFinesse, WS/7) that are used to drive the $5\text{s}5\text{p}~\ce{^{3}P_{0,1,2}}\leftrightarrow5\text{s}6\text{s}~\ce{^{3}S_{1}}$ transitions.

We further divide the red laser beam into three red MOT beams and three red cooling beams. The vertical and horizontal MOT beams are angled at $65^\circ$ with respect to the vertical axis of the glass cell to pass aside two microscope objectives mounted vertically, whereas the transverse MOT beams are aligned with the strong axis of the magnetic field gradient. The red cooling beams are oriented along the radial (R1, R2) and axial (A) directions. The two orthogonal radial cooling beams are angled at $45^\circ$ with respect to the transverse axis of the glass cell. The axial cooling beam is focused at the back aperture of the bottom objective to make it collimated at the output of the objective. 

We cool atoms in a 3D MOT operating first on the $\ce{^{1}S_{0}}\leftrightarrow\ce{^{1}P_{1}}$ broad dipole-allowed blue transition ($\lambda=460.9~\text{nm}$, $\Gamma/2\pi=30.2~\text{MHz}$) and then on the $\ce{^{1}S_{0}}\leftrightarrow\ce{^{3}P_{1}}$ narrow spin-forbidden red transition ($\lambda=689.5~\text{nm}$, $\Gamma/2\pi=7.4~\text{kHz}$). We create a blue MOT of $50\times10^6$~atoms at a temperature of a few mK, which we then transfer to a red MOT of roughly $10^6$ atoms at a temperature of $1.5~\mu\text{K}$.  The two pairs of three counter-propagating blue and red MOT beams are overlapped with dichroic mirrors. 


We calibrate the free-space resonance frequency of the 7.4~kHz $\ce{^{1}S_{0}}\leftrightarrow\ce{^{3}P_{1}}$ transition by performing excitation-depletion spectroscopy on the red MOT~(see Fig.~\ref{FigSidebandCooling}b). We use an excitation-depletion cycle composed of a $689~\text{nm}$ excitation pulse of $40~\mu\text{s}$ and a $688~\text{nm}$ depletion pulse of $10~\mu\text{s}$. We repeat this cycle up to five times to increase the depletion fraction, without significantly disturbing the resonance feature. By scanning the frequency of the excitation pulse in the low saturation regime, we determine the free space resonance with statistical error at the kHz level. We also use this technique to cancel stray magnetic fields by minimizing the Zeeman splitting observed in this feature.

We create two-dimensional arrays of optical tweezers using two acousto-optic deflectors (AA Opto-Electronic, DTSX-400-515) driven by polychromatic RF waveforms produced by two independent channels of an arbitrary waveform generator (Spectrum Instrumentation Corp., M4i6622-x8). We use a series of one-to-one telescopes ($f=300~\text{mm}$) to image the first AOD onto the second AOD and then the second AOD onto the back aperture of the bottom microscope objective. We stabilize the intensity of a single tweezer by monitoring the optical power after the first AOD and feeding back the output signal of a servo controller (New Focus, LB1005) into a voltage-variable attenuator (VVA) modulating the amplitude of the RF signal driving the first AOD. We use the same VVA to vary the trap depth of the tweezer.
We image atoms by scattering photons on the $\ce{^{1}S_{0}}\leftrightarrow\ce{^{1}P_{1}}$ transition with a transverse imaging beam oriented in the radial plane of the tweezer. The imaging beam is not retro-reflected to avoid standing waves or polarization gradients. We collect photons scattered by the atoms using two microscope objectives. The bottom objective, which is also used for focusing tweezers, images the scattered photons on a single-photon sensitive EMCCD camera (ANDOR, iXon 888), while the top objective collects additional photons that are retro-reflected back through the bottom objective to increase the photon collection efficiency.

We perform Sisyphus cooling and resolved sideband cooling using a combination of the four possible beam paths of the red laser: red MOT beams, radial cooling beams (R1, R2) and axial cooling beams (A). Although cooling can be achieved using several different beam geometries, we typically use the red MOT beams which allow us to cool in 3D and provide essentially all polarization components; however, retro-reflected cooling beams are not required for either Sisyphus cooling or sideband cooling. In particular, effective Sisyphus cooling is possible with only a single beam.  

\vspace{-3mm}
\section{Parity projection}~\label{appendix:parity_projection}
We prepare single atoms in tweezers using parity projection (PP). The initial number of atoms, $N$, loaded into the tweezer from the red MOT is assumed to follow a Poissonian distribution. This is projected onto a binary distribution by inducing pairwise loss between atoms in such a way that even values of $N$ are projected to $N=0$ and odd values of $N$ are projected to $N=1$. This approach to PP, which is ubiquitous in experiments with alkali atoms such as quantum gas microscopes~\cite{Bakr2010,Sherson2010} and tweezers~\cite{Schlosser:2001}, is induced by photo-association (PA) via diatomic molecular resonances~\cite{Jones2006}. Such molecular resonances have been identified for strontium in the electronically excited molecular potential which asymptotically corresponds to the $\ce{^{3}P_{1}}$~state~\cite{Zelevinsky2006, Reinaudi2012}. The first vibrational bound state in this potential has a binding energy of $-400~\text{kHz}$ with respect to the bare atomic resonance~\cite{Zelevinsky2006, Reinaudi2012}. 

\begin{figure}[t!]
	\includegraphics[width=\columnwidth]{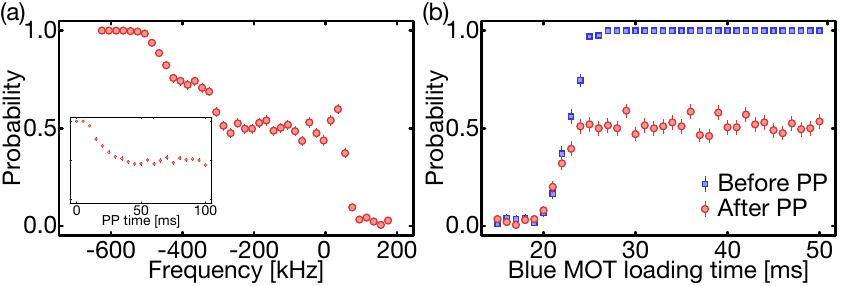}
	\caption{
	\textbf{Preparing single atoms via parity projection.}
	{(a)}~The probability of detecting an occupied tweezer after a parity projection (PP) pulse of 60~ms varies with the detuning of the addressing frequency. The highest step on the left corresponds to the situation when the PP pulse is detuned away from any atomic or molecular resonance, such that many atoms remain in the trap. The lowest step on the right corresponds to the heating on blue detuned motional sidebands, which expels atoms out of the trap. The plateau in the middle corresponds to the PP region where the occupation probability is $0.5$. {Inset:}~The probability of detecting an occupied tweezer monotonically decreases and saturates to $0.5$ as the duration of the PP pulse increases.
	{(b)}~As the blue MOT loading time increases, the initial number of atoms in the tweezer increases, such that the probability of detecting an occupied tweezer approaches $1.0$ before PP (blue squares), but saturates to $0.5$ after a PP pulse (red circles). The frequency detuning from free-space resonance here is $-226~\text{kHz}$.
	}
	\label{sFigParityProjection}
\end{figure}

We induce parity projection with a 60~ms excitation pulse on the $\ce{^{1}S_{0}}\leftrightarrow\ce{^{3}P_{1}}$ transition, detuned from the free-space resonance by $-226~\text{kHz}$~(Fig.~\ref{sFigParityProjection}a). The probability of detecting an occupied tweezer before PP is greater than $99.95\%$ for standard loading parameters, suggesting that numerous atoms are loaded into the trap on average. The occupation probability decreases and stabilizes to $0.5$ for a long PP pulse~(Fig.~\ref{sFigParityProjection}a, inset), characteristic of pairwise loss. Reliable single-atom preparation is further evidenced by the observation that the post-PP occupation probability of 0.5 is robust to the initial number of loaded atoms ~(Fig.~\ref{sFigParityProjection}b), which we can vary by loading our MOT for variable amounts of time, resulting in variable cloud densities.  

A quantitative understanding of the location and width of the PA feature is outside the scope of this work, but the resonance appears to lie between the binding energy of the molecular state at $-400~\text{kHz}$ and the red radial motional sideband of the atom in the trap at $-211~\text{kHz}$. We note that the internuclear separation of the molecular bound state in free space is $27~\text{nm}$~\cite{Zelevinsky2006} and may be reduced in the tweezer due to strong harmonic confinement. The Franck-Condon overlap between the bare atomic $\ce{^{1}S_{0}}$ state and the bound molecular state in $\ce{^{3}P_{1}}$ depends strongly on the internuclear separation between the atoms in the tweezer. This separation decreases as the atoms are cooled, so PA rates are possibly enhanced by cooling, thus skewing this feature closer to the red radial motional sideband.

\section{Fluorescence imaging}~\label{appendix:fluorescence_imaging}
\vspace{-10mm}
\subsection{Imaging fidelity}
We define imaging fidelity as the fraction of correctly identified images (a measure also known as classification accuracy).  An image is identified as either positive or negative by counting the number of photons detected in a certain region of interest and comparing this number to a fixed classification threshold.  We calculate fidelity by estimating the fraction of false positive and false negative identifications. These quantities are dependent on the choice of classification threshold, and different imaging conditions generally have different optimal choices of threshold.  For our quoted imaging fidelities in Fig.~\ref{FigFluorescenceImaging}b, we choose a fixed threshold for all times that is optimal for long times.  

False positives are readily estimated by measuring the number of false positives in a region of the image near the region onto which the atom is imaged.  We confirm that this nearby region produces the same number of false positives as the atomic region by also measuring the atomic region's false positives when atom loading is turned off.  

False negatives occur when an atom does not scatter enough photons to be detected.  This may happen because of two distinct reasons: (1) the imaging time was too short, or (2) the atom was lost before it could scatter enough photons.  False negatives due to (1) are estimated by fitting the single-atom histogram peak to a gaussian and computing the area of this fit that is below the classification threshold.  These types of false negatives tend to zero as imaging time is increased.  

Estimating type (2) false negatives requires knowledge about the loss mechanisms in play.  We show in the main text that we can reach regimes where losses are dominated by depopulation, such that the probability of loss is given by $p_s(N) = e^{-\chi \cdot N}$.  Having measured $\chi$, we estimate type (2) false negatives by integrating $\chi \cdot p_s(N)$ (properly normalized as a probability distribution) from zero up to the $N$ which corresponds to our classification threshold.  These false negatives depend only on the location of the threshold and are independent of imaging time for sufficiently long times.  Therefore, in the regime of long imaging times such that type (1) false negatives are negligible, optimal imaging fidelity is achieved for a choice of threshold which is a balance between minimizing false positives (requiring higher threshold) and minimizing type (2) false negatives (requiring lower threshold).  If imaging were lossless, unity fidelity could be reached by imaging for a long time and setting the threshold sufficiently high.  

Finally, we note that imaging fidelity may be increased in post-processing by weighing the photons detected on each pixel by the relative weight of that pixel in the averaged point spread function. We use this technique in all our quoted fidelities.  

\vspace{-2.5mm}
\subsection{Collection efficiency and radiation pattern}
We estimate the number of scattered photons by counting photons detected on our camera and estimating the collection efficiency of our imaging system.  This estimate takes into account the 0.84~sr solid angle of our NA = 0.5 objective, the measured transmission through all optical elements (0.47), the quoted quantum efficiency of our camera (0.76 at 461~nm), and a calibration of the camera gain via characterization of dark images~\cite{Bergschneider2018}. 

A large systematic error remains from the radiation pattern of the fluorescing atom.  A naive guess is that it is a dipole pattern ($f(\theta) = \sin^2{(\theta)}$) oriented along the polarization of the imaging beam. In this case, the collection efficiency varies by up to a factor of 7.3 between a polarization in the radial plane (best case) and one along the tweezer axis (worst case). 

We observe a dependence of the collection efficiency on imaging polarization that is consistent with a dipole pattern, insofar as collection is maximal when polarization is in the radial plane and minimal when it is axial.  We find that radial polarization not only maximizes detected photons, but also minimizes loss per detected photon, confirming that it truly increases collection efficiency and not just the scattering rate.  


However, a complete analysis of the radiation pattern would require accounting for the projection of the imaging beam polarization onto the coordinate frame defined by the tweezer polarization and estimating the scattering rates to each of the three non-degenerate states of $\ce{^{1}P_{1}}$, each of which have different radiation patterns.  We forgo such an analysis and instead assume that the radiation pattern is in between spherically symmetric and a dipole pattern along the radial plane.  We argue that this is a reasonable assumption because our imaging polarization is in the radial plane and we have confirmed that this does produce the best collection efficiency. The collection efficiency of a radial dipole pattern is 1.4 times higher than that of a spherically symmetric pattern.  This factor is the dominant source of error for $\chi^{-1}$. 


\section{Sisyphus cooling}~\label{appendix:sisyphus_cooling}
We measure the energy distribution of the atom after Sisyphus cooling using the adiabatic rampdown approach~\cite{Alt2003, Tuchendler2008}. Specifically, we measure the probability of an atom to remain in the tweezer after adiabatically ramping down the tweezer depth from its nominal value $U_0$ to some target value $U\leq U_0$. The cumulative energy distribution of the atom before the ramp down, $F(E/U_0)$, is obtained from the survival probability of the atom in the trap, $p_s(U/U_0)$, after converting the trap depth $U/U_0$ to the initial energy of the atom $E/U_0$ using the conservation of action argument~\cite{Tuchendler2008}. The mean energy of the atom is computed by integrating the cumulative energy distribution.

\begin{figure}[t!]
	\includegraphics[width=\columnwidth]{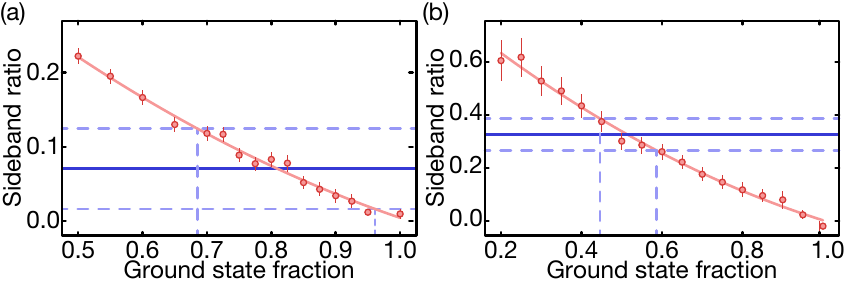}
	\caption{
	\textbf{Sideband thermometry.}
	{(a-b)}~Ratio of red to blue sideband amplitude as a function of ground state fraction, obtained via fitting simulated spectra for the (a) radial and (b) axial spectra. The dependence on ground state fraction is fitted to a quadratic function (red curve). The solid blue line is the fitted sideband ratio for our experimental data, with dashed lines representing a $1\sigma$ confidence interval. We quote a range of consistent ground state fractions where this confidence interval intersects the fitted quadratic function.  
	}
	\label{sFigSidebandThermometry}
\end{figure}

\section{Sideband thermometry}~\label{appendix:sideband_thermometry}
Unlike Raman sideband transitions which can be coherently driven without decay \cite{Kaufman2012,Thompson2013}, sideband transitions via direct excitation to $\ce{^{3}P_{1}}$ have inherent decay.  This complicates analysis because probing the sideband spectrum is unavoidably perturbative.  Since probing on the red sideband cools while probing on the blue sideband heats, the measured spectrum exhibits exaggerated asymmetry, and a naive analysis would underestimate the temperature. 

We therefore fit our measured sideband spectra to numerical simulation in order to extract a ground state fraction.  We simulate a driven 1D quantum harmonic oscillator, with decay implemented via quantum jumps~\cite{Molmer1993}.  The Hilbert space is defined as a product space of 20 motional states and 2 electronic states ($|g\rangle$ and $|e\rangle$). The non-Hermitian effective Hamiltonian is given by
\begin{eqnarray}
\mathcal{H}_{eff} &=& \mathcal{H}_0 + \mathcal{H}_l + \mathcal{H}_{\Gamma}\\
\mathcal{H}_0 &=& \hbar\omega (a^{\dagger}a+ \tfrac{1}{2}) \\
\mathcal{H}_l &=&  - \hbar\delta |e\rangle \langle e| + \tfrac{1}{2}\hbar\Omega (e^{i\eta(a + a^{\dagger})}|e\rangle \langle g| + \text{h.c.}) \\
\mathcal{H}_{\Gamma} &=& -i\tfrac{1}{2}\hbar\Gamma,
\end{eqnarray}
where $\omega$ is the angular trap frequency, $\delta$ is the detuning, $\Omega$ is the Rabi frequency, $\eta$ is the Lamb-Dicke parameter, and $\Gamma = 2\pi\times 7.4~\text{kHz}$ is the decay rate of the $\ce{^{3}P_{1}}$ state.  

The simulation proceeds in $\Delta t = 1~\mu$s timesteps.  At each timestep, the evolution operator $e^{-\tfrac{i}{\hbar}H_{eff}\Delta t}$ is applied to the state $|\psi\rangle$. $|\psi\rangle$ is then normalized and the probability of a quantum jump is computed as $p_{QJ} = p_e \Gamma \Delta t$, where $p_e = |\langle e|\psi\rangle|^2$ is the excited state population.  A quantum jump applies the operator $e^{i\vec{k}\cdot\hat{\vec{x}}}|g\rangle \langle e|$ to $|\psi\rangle$, where $\vec{k}$ is the wavevector corresponding to 689~nm light in a direction sampled from a dipole pattern.  Although the quantum jump operator is defined in 3 real dimensions, only its projection onto the relevant dimension is used.  

We run this simulation up to the same amount of time (74 $\mu$s) used for the probe in experiment.  In experiment, we use 3 such probe cycles, where at the end of each we use the 688~nm transition to project the electronic state to either the ground state or one of two $\ce{^{3}P_{J}}$ metastable states.  In simulation, this is implemented by running the probe cycle up to 3 times, where at the end of each cycle the quantum state is projected to the excited state with probability $\alpha\cdot p_e$, where $\alpha = 0.7$ is a projection fidelity factor which we find is necessary for a good fit to our data.  If the state is projected to the excited state, the simulation ends (representing loss, as measured in experiment).  If the state is projected to the ground state instead, the simulation either completes one more cycle or ends if 3 cycles have already been completed.  As there is also some baseline loss in our data, we implement this in post-simulation by projecting ground state populations to the excited state with probability given by our measured baseline loss.  

We compare the excited state population computed in simulation with the loss fraction measured in experiment.  As quantum jump is a stochastic method, we average over 2000 trials to obtain the final density matrix for each $\delta$ in our spectrum. The $\Omega$ used in simulation is chosen to fit the width of the carrier peak observed in experiment, and $\omega$ is chosen to fit the sideband frequency. 

We simulate spectra for various ground state fractions.  Ground state fraction is initialized by sampling the initial quantum state $|\psi (t=0) \rangle$ from a thermal distribution of motional eigenstates.  For each ground state fraction, we fit the amplitude of the red and blue sidebands and compute the ratio. We compare this to the ratio obtained by performing the same fit on our experimentally measured spectra, and find a range of ground state fractions for which our data is compatible with simulation (Fig.~\ref{sFigSidebandThermometry}).  
\bibliographystyle{h-physrev}
\bibliography{library2}

\end{document}